\documentstyle[psfig,epsfig]{mn}

\newif\ifAMStwofonts

\ifoldfss
  \ifCUPmtlplainloaded \else
    \NewTextAlphabet{textbfit} {cmbxti10} {}
    \NewTextAlphabet{textbfss} {cmssbx10} {}
    \NewMathAlphabet{mathbfit} {cmbxti10} {} 
    \NewMathAlphabet{mathbfss} {cmssbx10} {} 
  \fi
  \ifAMStwofonts
    \ifCUPmtlplainloaded \else
      \NewSymbolFont{upmath} {eurm10}
      \NewSymbolFont{AMSa} {msam10}
      \NewMathSymbol{\upi}     {0}{upmath}{19}
      \NewMathSymbol{\umu}     {0}{upmath}{16}
      \NewMathSymbol{\upartial}{0}{upmath}{40}
      \NewMathSymbol{\leqslant}{3}{AMSa}{36}
      \NewMathSymbol{\geqslant}{3}{AMSa}{3E}

    \fi
  \fi
\fi 
\newcommand{\be}{\begin{equation}}
\newcommand{\ee}{\end{equation}}
\newcommand{\ba}{\begin{eqnarray}}
\newcommand{\ea}{\end{eqnarray}}
\newcommand{\nn}{\nonumber \\}
\newcommand{\Mpc}{\, h^{-1}\!{\rm Mpc}}
\newcommand{\kms}{\, {\rm km}s^{-1}}
\newcommand{\r}{\mbox{\boldmath $r$}}

\newcommand{\tran}{{\!\perp}}
\newcommand{\thetab}{\mbox{\boldmath $\theta$}}
\newcommand{\de}{\partial}

\newcommand{\lgl}{\langle}
\newcommand{\rgl}{\rangle}

\newcommand{\x}{\mbox{\boldmath $x$}}
\newcommand{\k}{\mbox{\boldmath $k$}}
\newcommand{\Sb}{\mbox{\boldmath $S$}}
\newcommand{\N}{\mbox{\boldmath $N$}}
\newcommand{\Phib}{\mbox{\boldmath $\Phi$}}

\ifnfssone
  \newmathalphabet{\mathit}
  \addtoversion{normal}{\mathit}{cmr}{m}{it}
  \addtoversion{bold}{\mathit}{cmr}{bx}{it}
  \newmathalphabet{\mathbfit} 
  \addtoversion{normal}{\mathbfit}{cmr}{bx}{it}
  \addtoversion{bold}{\mathbfit}{cmr}{bx}{it}
  \newmathalphabet{\mathbfss} 
  \addtoversion{normal}{\mathbfss}{cmss}{bx}{n}
  \addtoversion{bold}{\mathbfss}{cmss}{bx}{n}
  \ifAMStwofonts
    \ifCUPmtlplainloaded \else
      \UseAMStwoboldmath
      \makeatletter
      \new@mathgroup\upmath@group
      \define@mathgroup\mv@normal\upmath@group{eur}{m}{n}
      \define@mathgroup\mv@bold\upmath@group{eur}{b}{n}
      \edef\UPM{\hexnumber\upmath@group}
      \new@mathgroup\amsa@group
      \define@mathgroup\mv@normal\amsa@group{msa}{m}{n}
      \define@mathgroup\mv@bold\amsa@group{msa}{m}{n}
      \edef\AMSa{\hexnumber\amsa@group}
      \makeatother
      \mathchardef\upi="0\UPM19
      \mathchardef\umu="0\UPM16
      \mathchardef\upartial="0\UPM40
      \mathchardef\leqslant="3\AMSa36
      \mathchardef\geqslant="3\AMSa3E
    \fi
  \fi
\fi 

\ifnfsstwo
  \DeclareMathAlphabet{\mathbfit}{OT1}{cmr}{bx}{it}
  \SetMathAlphabet\mathbfit{bold}{OT1}{cmr}{bx}{it}
  \DeclareMathAlphabet{\mathbfss}{OT1}{cmss}{bx}{n}
  \SetMathAlphabet\mathbfss{bold}{OT1}{cmss}{bx}{n}
  \ifAMStwofonts
    \ifCUPmtlplainloaded \else
      \DeclareSymbolFont{UPM}{U}{eur}{m}{n}
      \SetSymbolFont{UPM}{bold}{U}{eur}{b}{n}
      \DeclareSymbolFont{AMSa}{U}{msa}{m}{n}
      \DeclareMathSymbol{\upi}{0}{UPM}{"19}
      \DeclareMathSymbol{\umu}{0}{UPM}{"16}
      \DeclareMathSymbol{\upartial}{0}{UPM}{"40}
      \DeclareMathSymbol{\leqslant}{3}{AMSa}{"36}
      \DeclareMathSymbol{\geqslant}{3}{AMSa}{"3E}
    \fi
  \fi
\fi 

\ifCUPmtlplainloaded \else
  \ifAMStwofonts \else 
    \def\upi{\pi}
    \def\umu{\mu}
    \def\upartial{\partial}
  \fi
\fi

\title[Mapping the 3-D dark matter potential]{
Mapping the 3-D dark matter with weak lensing in~COMBO-17 }
\author[A.N. Taylor, et al] {
A.N. Taylor$^{1*}$, D.J. Bacon$^{1*}$, M.E. Gray$^{1*}$, C.
Wolf$^2$, K. Meisenheimer$^3$, S. Dye$^4$,\newauthor
 A. Borch$^3$, M. Kleinheinrich$^3$, Z. Kovacs$^3$, L. Wisotzki$^5$ \\
$^1$Institute for Astronomy, School of Physics, University of
Edinburgh, Royal Observatory, Blackford Hill, Edinburgh, EH9 3HJ,
U.K.\\
$^2$Department of Physics, University of Oxford, Keble Road,
Oxford OX1 3RH, U.K.\\
$^3$Max-Planck-Institut f\"{u}r Astronomie, K\"{o}nigstuhl 17,
D-69117, Heidelberg, Germany\\
$^4$Astrophysics Group, Blackett Laboratory, Imperial College,
Prince Consort Road, London SW7 2BW, U.K. \\
$^5$Astrophysikalisches Institut Potsdam, An der Sternwarte 16,
D-14482 Potsdam, Germany\\
$^*$email: ant@roe.ac.uk; djb@roe.ac.uk; meg@roe.ac.uk }
\date{}

\pagerange{\pageref{firstpage}--\pageref{lastpage}}

\pubyear{2003}

\begin{document}

\maketitle

\label{firstpage}

\begin{abstract}
We present a 3-dimensional lensing analysis of the $z=0.16$
supercluster A901/2, resulting in a 3-D map of the dark matter
distribution within a $3\times10^{5}[\Mpc]^3$ volume. This map is
generated from a combined catalogue of 3-D galaxy coordinates
together with shear estimates, using $R$-band imaging and
photometric redshifts from the COMBO-17 survey. To estimate the
3-D positions and masses of the main clusters in the supercluster
from lensing alone, we perform a $\chi^2$-fit of isothermal
spheres to the tangential shear pattern around each cluster as a
function of redshift. Motivated by the appearance of a second
cluster behind A902 in galaxy number density, we also fit a
two-cluster model to A902.

We then present the first 3-D map of the dark matter gravitational
potential field, $\Phi$, using the Kaiser-Squires (1993) and
Taylor (2001) inversion methods. These maps clearly show the
potential wells of the main supercluster components, including the
new cluster behind A902, and demonstrates the applicability of 3-D
dark matter mapping and projection free-mass-selected cluster
finding to current data. Finally, we develop the halo model of
dark matter and galaxy clustering and compare this with the auto-
and cross-correlation functions of the 3-D gravitational
potential, galaxy number densities and galaxy luminosity densities
measured in the A901/2 field. We find significant
anti-correlations between the gravitational potential field and
the galaxy number density and luminosities, as expected due to
baryonic infall into dark matter concentrations. We find good
agreement with the halo model for the number densities and
luminosity correlation functions, but some disagreement with the
shape of the gravitational potential correlation function, which
we attribute to finite-field effects.
\end{abstract}

\begin{keywords}
Gravitation;  gravitational lensing; Cosmology: observations, Dark Matter,
 Large-Scale Structure of Universe
\end{keywords}

\section{Introduction}

Gravitational lensing is a valuable method for measuring the
matter distribution in the Universe (see e.g. Mellier 1999,
Bartelmann \& Schneider 2001). Light rays are deflected by the
gravitational potential along their paths causing distortions in
the shapes of background galaxies together with a change in their
local number counts. These measurables are sensitive to all forms
of matter, whether visible or dark, which brings gravitational
lensing to the fore in studies of non-baryonic matter in the
Universe. For most regions of the sky, observations of shape
distortions (shear) reveal that the phenomenon is very weak
($\approx 1$\% change in ellipticity of an object). However we are
still able to measure this effect by averaging the ellipticity of
many galaxies to overcome the random intrinsic ellipticity of
galaxies.

Many studies have used weak shear measurements to measure the two
dimensional projected matter distribution of regions of space.
This has led to precise understanding of the masses and mass
profiles of galaxy clusters (see e.g. Tyson et al 1990, Kaiser \&
Squires 1993, Bonnet et al 1994, Squires et al 1996, Hoekstra et
al 1998, Luppino \& Kaiser 1997, Gray et al 2002). The 2-D shear
field is also used to measure the large-scale structure
distribution, and conseqeuently to measure cosmological parameters
(see e.g. van Waerbeke et al 2001, Hoekstra 2002, Bacon et al
2002, Refregier et al 2002, Jarvis et al 2003, Brown et al 2003,
Pen et al 2003).

Recently, there has been much interest in using redshift
information with weak gravitational lensing in order to add depth
to our picture of the mass distribution in the Universe. Lensing
tomography has been developed as a means of studying the growth of
the weak lensing power spectrum as a function of redshift (e.g.
Seljak 1998, Hu 1999, 2002, Huterer 2002, King \& Schneider
2002b), and the use of redshift information to remove the impact
of intrinsic galaxy alignments on lensing measurements has been
explored (Heymans \& Heavens 2002, King \& Schneider 2002a,b,
Heymans et al 2003). Observationally, Wittman et al (2001, 2002)
have used 3-D shear information to directly measure the mass and
redshift of two clusters discovered through this shear field.

A further significant development has been the realisation that a
full reconstruction of the 3-D gravitational potential and dark
matter density field is possible using just weak shear and
redshifts (Taylor 2001). A 3-D gravitational lensing analysis is
clearly of importance for cosmology. With a 3-D analysis we will
have, for the first time, a way of imaging the full dark matter
distribution from cluster scales upwards and over a order of
magnitude in redshift, independent of the galaxy distribution.
With such depth one can hope to visually see the growth of
structure. A 3-D lensing survey can also be used to construct a
mass-selected galaxy cluster catalogue, which would have many
important cosmological uses. A three dimensional lensing survey
would remove the projection effects that contaminate standard 2-D
projected surveys (White, van Waerbeke \& Mackey, 2002). In the
case of individual clusters this can cause biases in mass
estimation due to both foreground and background structures. With
a 3-D lensing analysis, both the mass and position of clusters can
be found independently of the baryonic content.

Following Taylor (2001), Hu \& Keeton (2003) subsequently
developed a pixelised version of the 3-D reconstruction, while
Bacon \& Taylor (2003) examined the practical implementation of
Taylor's (2001) method. Bacon \& Taylor (2003) demonstrated that
one can hope to reconstruct the gravitational potential on cluster
mass scales with ground-based shear surveys to a depth of $z=1$ by
including photometric redshifts with accuracy $\Delta z \simeq
0.05$.

To date the idea of reconstructing the full 3-D gravitational or
matter distribution has only been applied to simulated data.
However, data of this quality already exist; the COMBO-17 survey
(Wolf et al 2001) is a 17-band photometric redshift survey with
accuracy  $\Delta z = 0.05$ throughout $0<z<0.8$ and has already
been used for weak shear studies of a supercluster (Gray et al
2002), large-scale structure (Brown et al 2002), star formation
efficiency (Gray et al 2003) and galaxy alignment effects (Heymans
et al 2003). It is therefore an ideal dataset with which to carry
out a first direct reconstruction of the 3-D gravitational
potential.

In this paper we carry out a full 3-dimensional analysis of the
A901/2 supercluster field from the COMBO-17 survey, including
measurements of the 3-D shear, lensing potential and gravitational
potential fields. We describe the necessary methodological tools
for these measurements in Section 2. In particular we summarise
our approach to measurement of tangential shear, the
Kaiser-Squires inversion for the lensing potential and
reconstruction of the gravitational potential. In Section 2 we
also develop the halo model of mass and galaxy clustering allowing
an analysis of the correlations between the 3-D gravitational
potential and the galaxy number and luminosity densities. In
Section 3 we describe the COMBO-17 dataset, discussing the imaging
and photometric redshift data, and the means of obtaining a shear
catalogue. We describe the A901/2 field in detail, recalling the
relevant shear measurements obtained by Gray et al (2002) for this
supercluster.

In Section 4 we measure the 3-D tangential shear of the COMBO-17
A901/2 supercluster field, where we fit 3-D shear models to the
clusters, following the approach of Wittman et al (2001, 2002),
and measure the mass and redshift for the clusters directly from
the gravitational distortions. We examine the evidence for a
lensing signature of a cluster behind the supercluster at
$z=0.48$, and constrain its mass using a two-cluster fit to the
shear data. This is compared to the results of a 2-D lensing
analysis by Gray et al (2002). In Section 5 we calculate the
3-dimensional lensing potential and reconstruct the 3-dimensional
dark matter potential of the A901/2 supercluster field. We find
significant potential wells at the expected position of the
supercluster in 3-D, and also find a significant potential well at
the position of the background cluster. We discuss the measurement
significance for each of these detections.

In Section 6 we explore a natural application of 3-D gravity
measurements to measure the auto- and cross-correlation of the 3-D
gravitational potential with two other quantities: the galaxy
number density, and the luminosity distribution. We find that
there are anti-correlations for each of these quantities, and
describe how this can be understood in terms of baryonic infall
into dark matter concentrations. We compare our results with
theoretical predictions of the halo model. Finally, we present our
conclusions in Section 7.

\section{Methodology}

In this section we discuss the methodology we will use to extract
3-D information from our survey. We begin by discussing the
tangential shear, which we will use for a $\chi^2$ analysis of the
mass and position of mass concentrations in our survey. Following
this, we will describe our method for fully reconstructing the
gravitational potential in 3-D. Finally we present and develop the
halo model of nonlinear mass and galaxy clustering for analysing
correlation functions of the gravitational potential, the galaxy
number density and the galaxy luminosity fields.

\subsection{Tangential shear}\label{sec:gammat}

Weak gravitational shear distorts images in a way that can be
described by the shear matrix

 \be
    \gamma_{ij} =\left( \begin{array}{cc} \gamma_1 & \gamma_2 \\
                                 \gamma_2 & -\gamma_1 \end{array}
                                                        \right),
 \ee
where $\gamma_1$ and $\gamma_2$ represent the two orthogonal
states of distortion (see e.g. Bartelmann \& Schneider 2001). In
Section 4 we use the methods discussed by Kaiser, Squires \&
Broadhurst (1995) to obtain $\gamma_{ij}$ estimates for all
galaxies in our survey, and use the tangential shear measurements
to find estimates of cluster mass and redshift position (c.f. Gray
et al 2002). In order to do this, we choose a centre for each
cluster, which we take to be the optical centre (see Section 4).
We then calculate the tangential shear in radial bins around each
cluster,
\begin{equation}
    \gamma_t = -(\gamma_1 \cos 2 \theta + \gamma_2 \sin
    2\theta),
\end{equation}
where $\theta$ is the position angle taken from the cluster
centre. In order to estimate the mass and redshift of each
cluster, we will fit a singular isothermal sphere to the
tangential shear values (see Section 4).

\subsection{Lensing Potential}

To measure the 3-dimensional gravitational potential, we must
first calculate the lensing potential, $\phi$, which we consider
as a 3-dimensional field (c.f. Taylor 2001, Bacon \& Taylor 2003).
This is related to the shear field (in the limit of a flat sky) by
\begin{equation}
\gamma_{ij}(r,r\thetab) = \left(\de_i \de_j - \frac{1}{2}
\delta^K_{ij} \de^2\right)\phi (r,r\thetab), \label{gij}
\end{equation}
where $\thetab$ is a position angle on the sky,
 $\de_i \equiv r (\delta_{ij}- \theta_i \theta_j) \nabla_j = r(\nabla_i
- \theta_i \de_r)$  is the dimensionless transverse differential
operator and $\de^2 \equiv \de_i \de^i$ is the transverse
Laplacian. Here $r$ is a comoving distance,
 \be
    r(z) = \int_0^z \! \frac{d z'}{H(z')},
 \ee
 where
 \be
    H(z) = H_0 [\,\Omega_m (1+z)^3 + \Omega_v]^{1/2}
 \ee
 is the Hubble parameter, $\Omega_m$ is the present-day matter
 density parameter and $\Omega_v$ is the energy density associated
 with the vacuum. Throughout we shall assume a spatially flat universe
 with  $\Omega_m=0.3$ and $\Omega_v=0.7$.

The inverse relation, for calculating $\phi$ from $\gamma_{ij}$,
is given by the  Kaiser-Squires (1993) relation, generalised to
3-D:
\begin{equation}
    \widehat{\phi}(r,r\thetab) = 2 \de^{-4} \de_i \de_j \ \gamma_{ij}
(r,r\thetab),
\label{kaiser-squires}
\end{equation}
where $\widehat{\phi}$ is an estimate of $\phi$. In order to
calculate this quantity in practice, the shear field must be
smoothed in the transverse direction in order to overcome the
formally infinite noise amplitude (Kaiser \& Squires 1993). In
this equation we have used the operator $\de^{-2}$; this is the
(flat sky) inverse 2-D Laplacian operator, given by \be
        \de^{-2} \equiv \frac{1}{2 \pi}
    \int \! d^2 \!\theta \, \ln |\thetab - \thetab'|.
\ee
We measure transverse positions $\thetab$ in units of radians, leading
to dimensionless lensing quantities.

Equation (\ref{kaiser-squires}) reconstructs the lensing potential
up to an arbitrary function of comoving distance (Taylor 2001,
Bacon \& Taylor 2003):
\begin{equation}
    \widehat{\phi}(r,r\thetab) = \phi(r,r\thetab) + \psi(r,\thetab),
\end{equation}
  where
 \be
    \psi(r,\thetab) = \omega(r) +  \eta(r) \theta_x+\mu(r)\theta_y
    + \nu(r)(\theta_x^2 + \theta_y^2) ,
\label{unbiaspotom}
 \ee
and where $\omega(r)$, $\eta(r)$, $\mu(r)$ and $\nu(r)$ are
arbitrary functions. Here $\phi$ is the true lensing potential,
and $\psi(r,\thetab)$ is a solution to
 \be \left(\de_i \de_j - \frac{1}{2} \delta^K_{ij} \de^2 \right)
 \psi (r,\thetab)=0 .
 \ee
This arbitrary behaviour along the line-of-sight is due to the
fact that the shear only defines the lensing potential up to a
quadratic in angle for each slice in depth, and is the potential
analogue of the more familiar sheet-mass degeneracy for the lens
convergence, $\kappa$; $\phi = 2 \de^{-2} \kappa + C$. These terms
can be removed by taking moments of the measured lens
 potential over the area of a survey (Bacon \& Taylor 2003);
\be
    \psi_{mn}(r) = \frac{1}{A} \int_A \! d^2\theta \,
    \widehat{\phi}(r,r\thetab)
    (\theta_x^m + \theta_y^n),
    \label{psimn}
\ee
 where $A$ is the area of the survey. An estimate of the lensing
 potential with zero mean, gradient and paraboloid contributions
 is then given by
 \be
    \Delta\phi(r,r\thetab) = \widehat{\phi}(r,r\thetab)-
    \widehat{\psi}(r,r\thetab),
 \label{dp}
 \ee
 where $\widehat{\psi}$ is a correction term composed of the
 coefficients $\psi_{mn}$.
We have neglected higher order polynomial
contributions to $\phi$ which could in general lead to a bias in
the estimated gradient and paraboloid terms; however, in practice
our boundary conditions at the field edge (continuity for
amplitude and gradient) mean that we do not have such terms in our
reconstruction.

For a general survey geometry, the coefficients $\psi_{mn}$ can be
calculated numerically. The removal of the first few moments will
reduce the power on the largest scales of the survey. This effect
and the noise properties of the reconstructed $\Delta \phi$ field
are discussed further in Bacon \& Taylor (2003).

\subsection{Gravitational Potential}

Having calculated an estimate of the lensing potential, $\Delta
\phi$, we are now in a position to reconstruct the 3-D
gravitational potential. The lensing potential is related to the
gravitational potential by the line-of-sight integral
\begin{equation}
\phi(r,r\thetab) = 2 \int_0^{r}\!\! dr' \, \left(\frac{r-r'}{r
r'}\right) \Phi(r',r'\thetab), \label{pot}
\end{equation}
where we have assumed a spatially flat universe and the Born
approximation. Treating $\phi(r,r\thetab)$ as a 3-D variable, we
can invert equation (\ref{pot}) to arrive at an estimate of the
3-D gravitational potential (Taylor 2001):
\begin{equation}
\Phi(r,r\thetab) = \frac{1}{2} \de_r r^2 \de_r \,
\phi(r,r\thetab), \label{inv}
\end{equation}
where $\de_r = \thetab.\nabla$ is a radial derivative. The
solution to equation (\ref{inv}) can be checked by direct
substitution into equation (\ref{pot}) and integrating by parts.
In practice we shall use the corrected potential, $\Delta \phi$,
from equation (\ref{dp}), yielding
 \be
    \Phi(r,r\thetab) =  \frac{1}{2} \de_r r^2
     \de_r \Delta \phi(r,r\thetab)
    \label{unbiaspot}
 \ee
 as an estimate of the dark matter potential. For completeness we
 note that the density field is
\be
    \delta(r,r\thetab) = \frac{ a(r)}{3 \Omega_m H_0^2 }
        \nabla^2 \de_r r^2
     \de_r \Delta\phi(r,r\thetab),
     \label{unbiasdel}
 \ee
 where $a$ is the expansion factor.
 However the dataset we shall consider here is too noisy to
 reconstruct the 3-D density field. Hence we shall use the
 set of equations, (\ref{kaiser-squires}), (\ref{psimn}),
(\ref{dp}) and (\ref{unbiaspot}) to generate the 3-D lensing
convergence, 3-D lensing potential and 3-D dark matter potential
from combined shear and redshift information.

\subsection{Wiener filtering}

Hu \& Keeton (2003) and Bacon \& Taylor (2003) have shown that
reconstruction of the 3-D gravitational potential benefits
enormously from Wiener filtering. For realistic lensing surveys
cluster-mass concentrations are typically undetectable in the
unfiltered gravitational potential, due to the large shot-noise.
However cluster masses can be detected with a signal-to-noise
ratio of $\simeq 2$ \emph{per pixel} in the Wiener filtered field.
We need only filter in the redshift direction (Bacon \& Taylor,
2003), leaving the potential field in the transverse direction
unfiltered.

For each $x, y$ position, we construct a vector, $\Phi_i$,
composed of the gravitational potential measurement in the
$i^{th}$ redshift bin, along the $z$ direction. We also construct
a noise covariance matrix, $N_{ij}$, for the gravitational
potential covariance in the $z$ direction, measured from 100
reconstructions of the $\Phi$-field with randomised ellipticities.
A further matrix, $S_{ij}$, represents the expected signal
covariance along the line of sight. For this we use a unit matrix
multiplied by $S=(1.3\times 10^{-7})^2$ which equals the square of
the expected gravitational potential amplitude for a $5\times10^{13}
M_\odot$ cluster (c.f. Bacon \& Taylor 2003);
 \be
        S_{ij} = S \delta_{ij}.
 \ee
  We then apply the Wiener filtering,
\begin{equation}
\label{eqn:wiener} \Phib' = \Sb (\Sb+\N)^{-1} \Phib,
\end{equation}
where $\Phib'$ is the Wiener filtered gravitational potential we
require.

\subsection{Statistical properties and the halo model}
\label{halo}

Having shown how one can reconstruct the 3-D Newtonian
gravitational potential from lensing, we now turn to the
statistical properties of the reconstructed field. In addition to
the gravitational field, we can also measure the 3-D galaxy number
density, $n(\r)$, and the luminosity density, $L(r) \equiv
\rho_L(\r)$. It is of interest to estimate the auto- and
cross-correlations of these fields from theory, in order to
compare with the results of our data analysis.

In practice we shall use a projected correlation function, where
the statistical correlations are calculated in slices in redshift
and then averaged. This is convenient since each redshift bin is
much larger than the transverse size of the field, and helps to
reduce the effect of the line-of-sight Wiener filter which will
distort the statistical properties of the filtered field. We
define the projected correlation between two fields, $X$ and $Y$,
by
 \be
    C^{XY}(r_\tran) = \frac{1}{N_z} \sum_{i=1}^{N_z}
        \langle X(\x_{\!\perp},z_i) Y(\x_{\!\perp}+\r_{\!\perp},z_i)
        \rangle,
        \label{corrln}
\ee
 where $X$ and $Y$ denote the fields ($\Phi$, $n$, $L$),
 $\x_{\!\perp}$ and $\r_{\!\perp}$ are transverse position vectors
 in the slice at redshift $z_i$, and $r_\tran=|\r_{\!\perp}|$.
 The number of redshift slices is $N_z$. The averaging,
 $\lgl \cdots \rgl$, is taken over the fields at each slice. In
 practice this averaging will be done over a finite survey on the
 sky with pixelised data.

The observed $\Phi$-$\Phi$, $\Phi$-$n$ and $\Phi$-$L$ projected
auto- and cross-correlation functions can be generated from their
corresponding power spectra by the transformation
 \be
    C^{XY}\!(r_\tran) =
    \int\!\! \frac{ k^2 dk }{2 \pi^2}\, W(k,r_\tran) P_{XY}(k)
    \left|\omega_{\rm pix}(k)\right|^2,
 \label{corr}
 \ee
 where for a cylinderical survey with radius $R_s=4 \Mpc$
 and length $L_s=1000 \Mpc$ the survey window function
 is given by
 \ba
    W(k,r_\tran) &=& \int_{-1}^{1} d \mu \,j_0^2(k L_{\rm pix} \mu)
 [ 1-W_\circ^2(k R_s\sqrt{1-\mu^2})] \nn
     &  \times &
        [J_0(k r_\tran \sqrt{1-\mu^2}) - W_{\circ} (k R_s\sqrt{1-\mu^2}) ]
        \label{pow-wind}
 \ea
 where
 \ba
    W_\circ(x) = \frac{2 J_1(x)}{x},
 \ea
 is the 2-D Fourier transform of a circular survey aperture.
$L_{\rm pix}=50 \Mpc$ is the pixel length in the redshift
direction. We also include the effects of pixelisation in the
transverse direction with the pixel window function;
 \be
    \omega_{\rm pix}(k) = \int_{-1}^{+1} \! d\mu \,
        W_\circ (k R_{\rm pix} \sqrt{1-\mu^2}) ,
 \ee
for circular pixels of radius $R_{\rm pix}=0.2\Mpc$. The various
terms in equation (\ref{pow-wind}) then correct for pixelisation,
the survey window function, and removal of the mean field. There
is also an integral constraint, that the integral of the field is
zero when averaged over the survey volume.

The statistical properties of dark matter and galaxies can be most
easily modeled theoretically with the halo model (e.g. Peacock \&
Smith 2000, Seljak 2000). Here the statistical distribution of
dark matter is considered as a sum of linear correlations between
haloes on large scales and internal correlations within haloes on
small scales. These haloes can be populated with a distribution of
galaxies and allocated a luminosity, depending on the halo mass.
The statistical properties of the haloes and galaxies can then be
calculated by averaging over the mass function. The power spectra
we require is given by a linear and a nonlinear term;
 \be
    P_{XY}(k) = \int \frac{d n(M)}{(2 \pi)^3}
    \left( \frac{\rho(k,M)}{\bar{\rho}}\right)^2W_{XY}(k,M) +
    P^{\rm LIN}_{XY}(k),
    \label{powerXY}
 \ee
 where $dn(M) = dM (dn/dM)=(\bar{\rho}/M) f(\nu) d \nu$ and $f(\nu)$
 is the Tolman-Sheth mass function,
 \be
    \nu f(\nu) = A \left(1+ \left(\frac{\nu}{\sqrt{2}}\right)^{-0.3} \right)
    \left( \frac{\nu}{\sqrt{2}}\right)^{1/2}
                    e^{-\nu/2\sqrt{2}},
 \ee
 where $\nu(M,z) = (\delta_c(z)/\sigma(M,z))^2$ and $\sigma^2(M,z)$
 is the variance on the scale of haloes of mass $m$ at redshift $z$.
 We use a critical
 collapse overdensity of $\delta_c=1.68/(1+z)$. The halo profile,
 $\rho(r,M)$, is assumed to be a Navarro-Frenk-White (NFW; 2000)
 profile;
 \be
    \rho(r,M) \propto  \frac{1}{(r/r_s)^{-\alpha} (1+
    r/r_s)^{3+\alpha}},
 \ee
 where the inner slope is $\alpha=-1$, the inner core radius is
 $r_s=r_v/c$, where $r_v=3M/(4 \pi \bar{\rho} \delta_{\rm vir})$ is
 the virial radius and $c=10(M/M_*)^\beta$ is a halo concentration factor. He we assume
 $M_*=8 \times 10^{11} h^{-1}{\rm M}_\odot$ (Guzik \& Seljak 2002) and
 $\beta=-0.2$. The Fourier transform of the NFW halo profile with a
 sharp cut-off at the virial radius is given by
 \ba
    \rho(k,M) &= & k \cos (k) [{\rm Ci}(k(1+c)) - {\rm Ci}(k)] + \nn
            & &
                k \sin (k) [{\rm Si}(k(1+c)) - {\rm Si}(k)]
                - \frac{\sin ck}{1+c},
 \ea
 where Ci$(z)$ and Si$(z)$ are the Cosine and Sine integrals,
 respectively. For simplicity we shall assume that the linear power
 spectrum of all of the fields is just equal to the linear dark
 matter power spectrum.

 The weighting functions in the nonlinear, halo-halo part of
 equation (\ref{powerXY}) are given by
 \be
    W_{\Phi\Phi}(k,M) = \frac{9}{4} H_0^4 \Omega_m^2 (1+z)^2 k^{-4}
 \ee
 for the potential power spectrum, and
 \be
    W_{nn}(M) =
     \frac{\lgl N (N-1)\rgl \bar{\rho}^2}{(\bar{n}M)^2}
 \ee
 for the galaxy-galaxy power spectrum,
 where $\lgl N(N-1)(M) \rgl$ is the variance of the number of
 galaxies in a halo of mass $M$. The galaxy-mass weight is
 given by
 \be
    W_{n\Phi}(k,M) = - \frac{3}{2} H_0^2 \Omega_m (1+z) k^{-2}
    \frac{\lgl N \rgl \bar{\rho}}{\bar{n}M},
 \ee
 where $\lgl N(M) \rgl$ is the average number of galaxies for a
 halo of mass $M$.
 We model the halo occupation numbers with the same functional form;
 \be
    \lgl N \rgl, \, \, \lgl N(N-1) \rgl^{1/2}
    = \left(\frac{M}{M_*} \right)^{0.81}
    \left( 1 - e^{-(M/M_0)^2/2}\right),
 \ee
 where $M_*=10^{13.47} {\rm M}_\odot$. For the mean
 occupation number, $\lgl N \rgl$,  we use $M_0=10^{13}{\rm
 M}_\odot$, while for the scatter in the occupation number,
 $\lgl N(N-1)\rgl$,
 we use $M_0=6\times 10^{12}{\rm M}_\odot$. These parameters are
 suitable for all galaxies with an absolute magnitude threshold of
  $M_B<-19.5$ (eg Seljak 2000).

\begin{figure}
 \centerline{\psfig{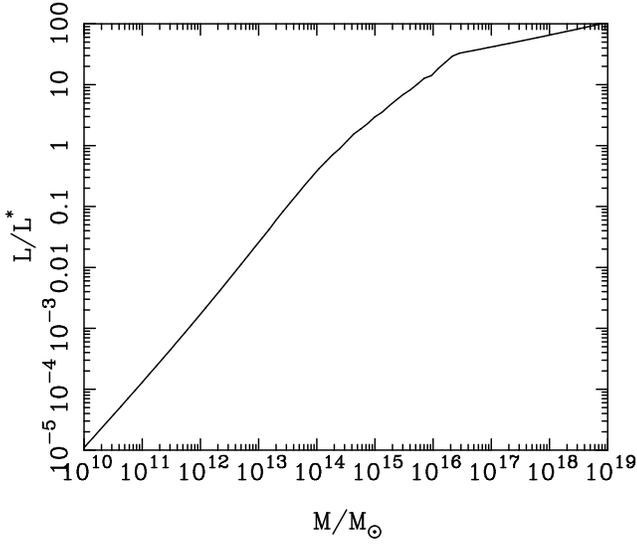}}
 \caption{The luminosity-mass relation, $L(M)$, for the COMBO-17 data set,
 estimated from equating the cumulative R-band luminosity and
 Tolman-Sheth mass functions. }
 \label{fig-LM}
\end{figure}

 The luminosity-mass weight is given by
 \be
    W_{L\Phi}(k,M) =- \frac{3}{2} H_0^2 \Omega_m (1+z) k^{-2}
    \left \lgl  \frac{M}{L} \right\rgl
     \frac{L(M)}{M},
 \ee
 where $\lgl M/L \rgl$ is the universal mass-to-light ratio
 and $L(M)$ is the average luminosity of a halo of mass $M$.
 Finally the luminosity-luminosity weight is
 \be
    W_{LL}(M)=
    \left \lgl  \frac{M}{L} \right\rgl^2
    \left( \frac{L(M)}{M} \right)^2.
 \ee
 The luminosity-mass relation, $L(M)$, can be estimated by
 equating the cumulative halo number density to the cumulative
 luminosity function, $N(>M) = N(>L)$ (Peacock \& Smith 2000).
 We use the R-band luminosity function derived by Wolf et al
 (2003) for the COMBO-17 data. Figure \ref{fig-LM} shows the
 derived luminosity-mass relation for the COMBO-17 dataset.

 We should also consider the range of the mass
 integration in equation (\ref{powerXY}). In a finite survey,
 the high-mass end of the mass function may not be
 well sampled, as high mass concentrations are rare. To account
 for this, and to compare results with a real survey, we may
 truncate this integral at the largest mass found in the survey.
 This will have the effect of suppressing the largest nonlinear
 scales. As the potential field is weighted towards larger scales,
 this can have an important effect on the statistics of the potential
 field. We discuss this further in Section 6.

 Having presented our methods for recovering the 3-D
 dark matter gravitational potential field, and the halo model for analysing the
 auto- and cross-correlation of the gravitational potential, number
 density and luminosity densities, we now turn to the
 dataset we will apply this analysis to.

\section{The COMBO-17 Data}\label{sec:obs}

\subsection{Observations}

The quality of data required for a full 3-D gravitational lensing
analysis already exists. The COMBO-17 survey is a 17-band
photometric redshift survey with gravitational lensing quality
R-band data (Wolf et al 2001). The survey consists of four
separate areas, each of $30' \times 30'$ comprising a total of 1
square degrees. All of the fields were observed using the
Wide-Field Imager (WFI) at the MPG/ESO 2.2m telescope on La Silla
in Chile, with a $4\times2$ array of $2048\times 4096$ pixel CCDs,
each pixel subtending 0.238 arcseconds. One of these fields is
centred on the A901/2 supercluster which has previously been
analysed by Gray et al (2002) and it is this field that we will
use in this analysis.

\subsubsection{Photometric redshifts}
Each of the COMBO-17 fields was observed in 17 different filters,
with the intention of obtaining object classification and accurate
photometric redshifts. In order to provide reliable redshifts, the
filter set included five broad-band filters ($UBVRI$) and 12
medium-band filters from 350 to 930 nm. This observing strategy
allows simultaneous estimates of Spectral Energy Distribution
(SED) classifications and photometric redshifts from
empirically-based templates.  Wolf et al (2001) describe in detail
the photometric redshift estimation methods used to obtain typical
accuracies of $\sigma_z=0.05$ for galaxies throughout $0<z<1$.

\subsubsection{Shear measurements}
The $R$ filter was used in best seeing conditions throughout the
observing campaign, in order to provide a deep $R$-band image from
which to measure the gravitational shear. Gray et al (2002)
discuss the procedure used to reduce the $R$ band imaging data,
which totalled 6.3 hours. As described there, the 352 individual
chip exposures for the A901/2 field were registered using linear
astrometric fits, with 3$\sigma$ bad pixel rejection to remove bad
columns and pixels.

We then applied the {\tt imcat} shear analysis package, using
Kaiser, Squires \& Broadhurst's (1995) weak lensing measurement
method, to our reduced image (see Gray et al 2002 for details).
This resulted in a catalogue of galaxies with centroids and shear
estimates throughout our field, corrected for the effects of PSF
circularisation and anisotropic smearing.  We appended to this
catalogue the photometric redshifts estimated for each galaxy from
the standard COMBO-17 analysis of the full multicolour dataset. Of
the 37,243 galaxies in the shear catalogue, 36\% had a reliable
photometric redshift, the remainder being fainter than the $R=24$
reliability limit of the redshift survey.  The requirement for the
3-D lensing study that the redshift of each galaxy be known
clearly results in an immediate reduction in available galaxies,
as it is apparent that most of the background sample is composed
of galaxies that are small and fainter than the magnitude limit of
the redshift survey.

\subsection{The A901/2 supercluster}

\begin{figure}
\psfig{figure=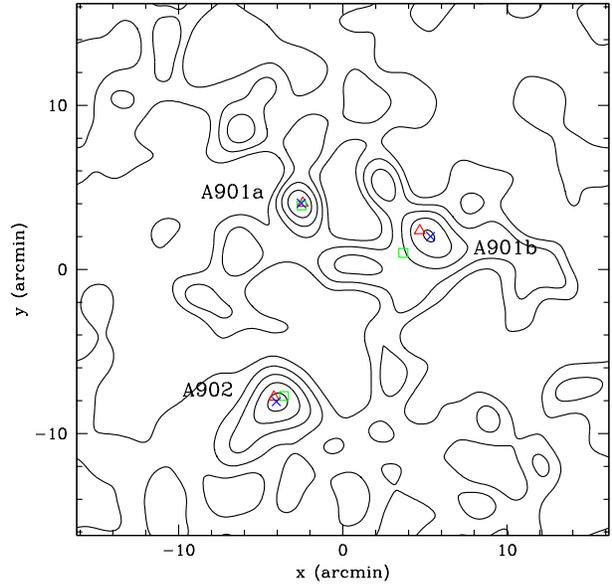,width=\columnwidth,angle=0}
\caption{Two-dimensional projected distribution of mass in the
A901/2 field.  The contours show the lensing mass map of Gray et
al. (2002), with crosses, squares, and triangles marking the
location of the peaks in the mass and light distributions and the
positions of the brightest cluster galaxies, respectively. The
contour levels are spaced by $\Delta \kappa =0.02$, while the
noise level is $\sigma_\kappa=0.027$. }\label{fig-massmap}
\end{figure}

The supercluster itself is composed of three clusters of galaxies
(A901a, A901b, and A902), all at $z=0.16$ and contained within the
$0.5^{\circ}\times 0.5^{\circ}$ field-of-view of the WFI. The
X-ray, number count, and 2-dimensional lensing properties of the
field are discussed in detail in Gray et al (2002). Figure
\ref{fig-massmap} shows the projected mass distribution estimated
by Gray et al (2002). The three main components of the
supercluster, A901a, A901b and A902 are clearly detected. One of
the main conclusions of that work was that the distribution of the
early-type supercluster galaxies (as selected by their location on
the $B-R$ vs $R$ colour-magnitude diagram) does not fully trace
the distribution of the lensing-revealed mass map.  In particular,
misalignments between (early-type) light and mass were found in
A901b which is the only cluster to display extended X-ray emission
in pointed ROSAT HRI observations by Schindler (2000).

Additionally, no single mass-to-light ratio (again, considering
only the light from the colour-selected early-type galaxies) could
be adequately determined for the system as a whole: A901a, by far
the most luminous of the clusters, was under-represented in the
lensing mass map.  Clearly, the mapping from early-type light to
mass in this system is not a linear one, either due to dynamical
effects from an ongoing merger or to divergent paths in galaxy
evolution (and hence total luminosity) in the three clusters.

The Gray et al. (2002) lensing study employed a standard weak
lensing approximation and assumed that all the mass measured in
the projected mass map was due to a mass concentration in a single
lens plane. However, with the advent of 3-dimensional mapping
techniques, we are now able to fully probe the mass distribution
throughout the entire volume of the COMBO-17 observations.  This
advance allows us to test whether the bulk of the mass does indeed
lie in the redshift plane of the supercluster at $z=0.16$, or
whether projection effects from matter behind the supercluster
contribute significantly. If projection effects are found to
contribute significantly to the 2-dimensional mass map, we may
have an alternative explanation to discrepant $M/L$ ratios
observed in the system.

\section{Measuring the Mass and 3-D Position of Clusters}

\subsection{Single cluster model}

\begin{figure*}
\psfig{figure=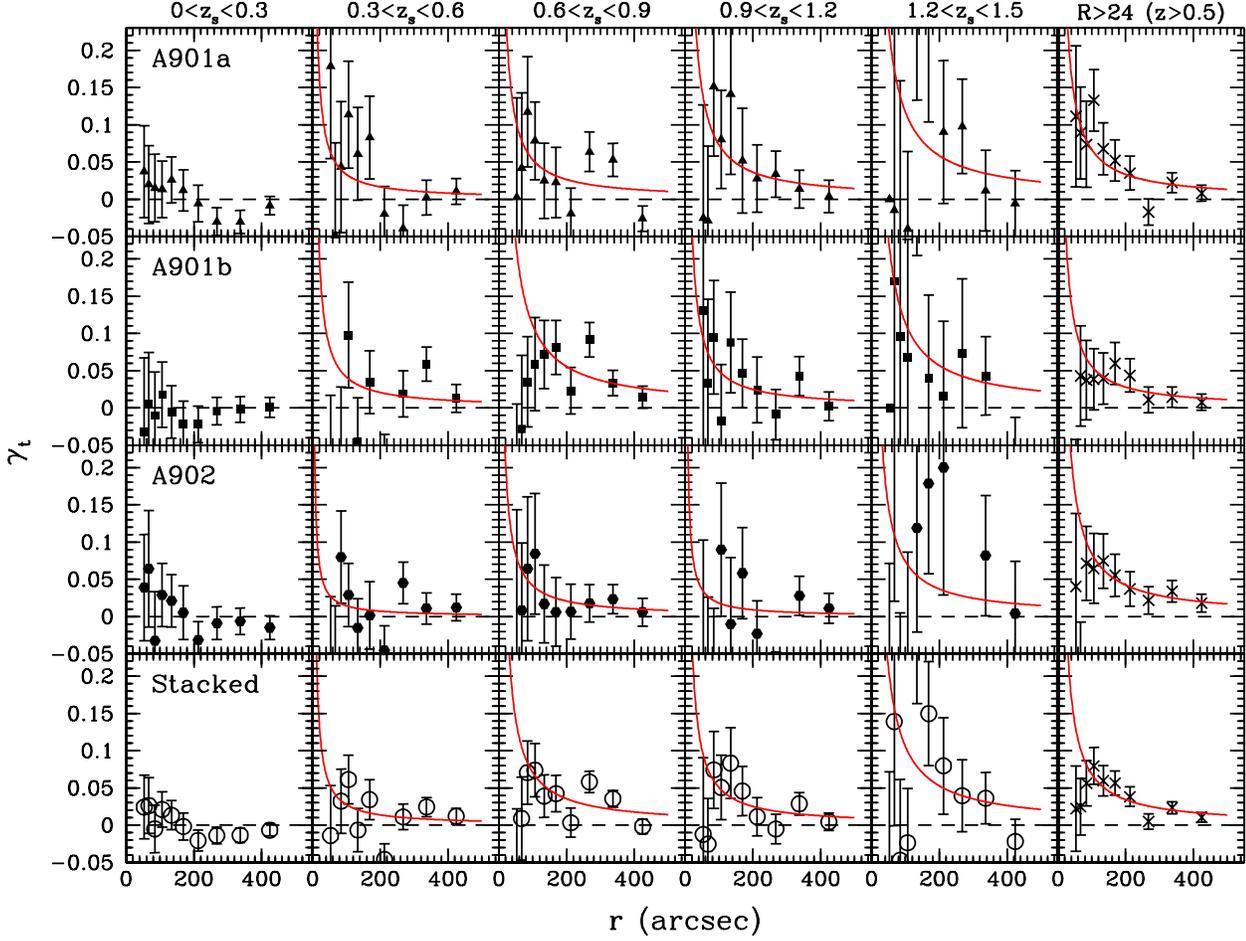,height=\textwidth,angle=270}
\caption{Tangential shear as a function of  angular radius and
redshift. Each row presents the radial profile of the tangential
shear around each of the three mass peaks in the field, according
to increasing source redshift.  The solid curve in each case
represents with best-fit SIS model for the redshift bin, with
$z_{\rm lens}$=0.16. Note the growth of the lensing signal as the
slices in redshift space become more distant. The bottom row shows
the signal for an `average' cluster, calculated by stacking the
catalogues for the three clusters around a common origin (no
attempt is made to scale the signal according to varying cluster
mass).} \label{fig-biggtzplot}
\end{figure*}

Using the shear estimates for the galaxies in the A901/2 field,
together with photometric redshifts, we are able to measure the
redshifts and masses of the clusters from the 3-D distortion field
that they create.  While constraining cluster masses has been the
traditional goal for weak lensing studies, it is also of
importance to be able to place additional constraints on the mass
distribution along the line of sight using a method that is
independent of baryonic content.  While in this case the
structures we are probing are clear overdensities of galaxies at
known redshift, we shall use this supercluster field as an
illustration of what can be achieved when the shear signal from
each lensed galaxy can be appropriately weighted by its redshift.
One possible application of such a technique would be as a way to
confirm and constrain the masses and redshift of `dark clusters'
(Erben et al. 2000, Gray et al. 2001, Umetsu \& Futamase 2003,
Weinberg \& Kamionkowski 2003), or to deproject multiple mass
concentrations along a single line-of-sight. In addition, 3-D
lensing offers the possibility of creating mass-selected samples
of dark matter concentrations.  Such a sample would allow for
cleaner comparisons with theories of structure formation, without
resorting to a priori assumptions about how the dark matter and
baryonic distribution (in the form of galaxies or hot X-ray gas)
are related.

Consequently, in the following analysis we will treat the masses and
(known) redshifts of each cluster as free parameters.  Figure
\ref{fig-biggtzplot} shows the radial profile of the tangential shear,
$\gamma_t(\theta)$ (Equation 2), as a function of source redshift for
each cluster, $z_s$. In this figure each row represents a different
cluster, with the final row representing the results from a `stacked'
cluster, obtained by recentering all three catalogues about a common
origin and concatenating. Each column shows the shear signal in
increasing source redshift slices. The solid curves show the best-fit
tangential shear for a singular isothermal sphere (SIS) model;
\begin{equation}
    \gamma_{t,{\rm SIS}}(\theta) = \frac{2\pi}{\theta}
    \left(\frac{\sigma_v}{c}\right)^2 \frac{D_{ls}}{D_s}.
\end{equation}
Here $D_{ls}$ and $D_s$ are the angular distances from lens to source
and from observer to source, respectively. We fix the lens redshift at
$z_{\rm lens}=0.16$ so that the SIS model is characterized only by a
single parameter, the halo velocity dispersion, $\sigma_v$. The final
column in Figure \ref{fig-biggtzplot} shows the shear profile for the
$\sim$65\% of source galaxies with $R>24$, for which photometric
redshifts are unreliable. As expected, the lensing signal generally
grows as a function of source redshift behind each cluster, and no
signal is detected in the lowest redshift ($0<z_s<0.3)$ bin where many
of the source galaxies are foreground to the lens.

\begin{figure}
\centering
\begin{picture}(200,260)
\includegraphics{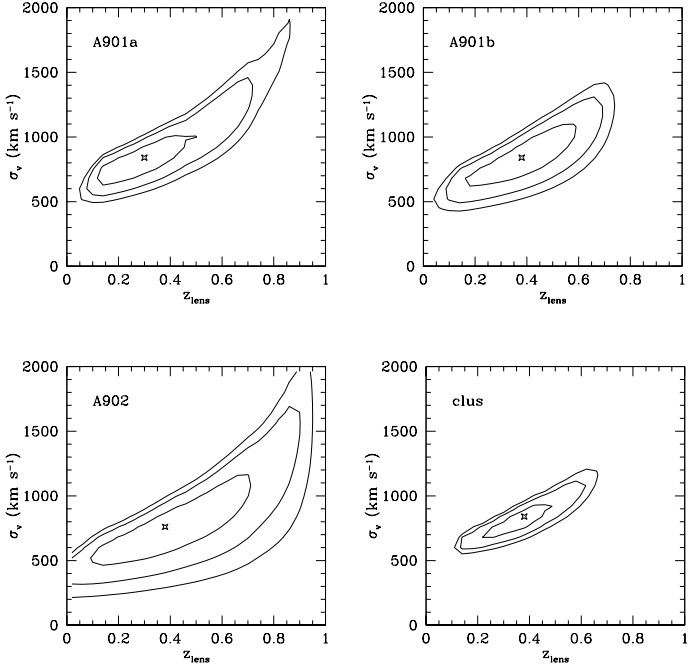}
\end{picture}
\caption{Constraints on cluster mass and redshift. Each panel
shows the 1, 2 and 3-sigma (68\%, 95\% and 99.5\% confidence)
contours for a $\chi^2$ two-parameter fit to the data from Figure
1, allowing both $z_{\rm lens}$ and $\sigma_v$ (for a SIS model)
to vary for each cluster.} \label{fig-gtfitcont}
\end{figure}

Figure \ref{fig-gtfitcont} shows the constraints on mass and
redshift for the clusters from a 2-parameter $\chi^2$ fit of the
data points from Figure \ref{fig-biggtzplot}, allowing both
$z_{\rm lens}$ and $\sigma_v$ to vary. For A901a the best-fit
$z_{\rm lens}$ is $z_{\rm lens}=0.30^{+0.10}_{-0.16}$ and a small
range in preferred mass of $\sigma_v=840^{+100}_{-105}\kms$. For
A901b, we obtain $z_{\rm lens}=0.38^{+0.14}_{-0.11}$ and
$\sigma_v=840^{+170}_{-110} \kms$. For A902 the best-fit
parameters are $z_{\rm lens}=0.38^{+0.23}_{-0.20}$ and
$\sigma_v=760^{+220}_{-200}\kms$.  As expected, the increased
number density in the `stacked' catalogue results in slightly
smaller error bars: $z_{\rm lens}=0.38^{+0.05}_{-0.09}$ and
$\sigma_v=840^{+25}_{-100}\kms$.  In each case the degeneracy in
the contours is clear and the best-fit redshift is biased towards
higher redshifts, although in no case is the true solution of
$z_{\rm lens}=0.16$ excluded at the 95\% confidence level.  This
is partially the result of the coarseness of the redshift bins:
much of the signal constraining the redshift of the cluster comes
from the rapid rise in lensing signal just beyond the lens
redshift.  In this case our lowest-redshift bin spans the
relatively wide redshift slice of $0<z_s<0.3$.  A finer redshift
slice would remove some of this degeneracy, for example if $\Delta
z=0.1$ the best-fit solution for the redshift of A901a falls to
$z_{\rm lens}=0.22$.  However, a balance needs to be struck
between ensuring fine adequate sampling in redshift space and
having sufficient numbers of galaxies in each redshift bin.

Note that in each case, there is a strong detection of the
clusters; the $\chi^2$ results are not consistent with zero mass.
This provides us with estimates of the detection significance of
the clusters by marginalising over redshift, and examining how
likely zero mass is in the resulting probability contours. For
A901a, we find that the probability of a zero-mass solution at
$z_{\rm lens}=0.16$ is $4.7\times10^{-8}$; for A901b,
$8\times10^{-6}$; and for A902, $9\times10^{-3}$.  Thus all
clusters are highly significant mass concentrations.

If we fix the redshift of the clusters at $z=0.16$ and fit only
the velocity dispersions, then the best-fit results are: A901a,
$\sigma_v=680^{+25}_{-90}\kms$; A901b,
$\sigma_v=600^{+40}_{-85}\kms$; and A902,
$\sigma_v=520^{+55}_{-90}\kms$. These values are our best
estimates for a single cluster model and are tabulated in Table 1,
along with the implied total mass within a $0.5 \Mpc$ radius.
These results can be compared with those of Gray et al (2002) who
measured the tangential shear for the faint $R>22$ galaxies
without any redshift information. They found the velocity
dispersion of $\sigma_v = 542^{+195}_{-333}\kms$ for A901a,
$\sigma_v = 659^{+129}_{-161}\kms$ for A901b and $\sigma_v =
738^{+244}_{-384}\kms$ for A902. All of these are in consistent
within the errors with our current measurements for a similar
fixed redshift of $z=0.16$.

\begin{figure*}
\psfig{figure=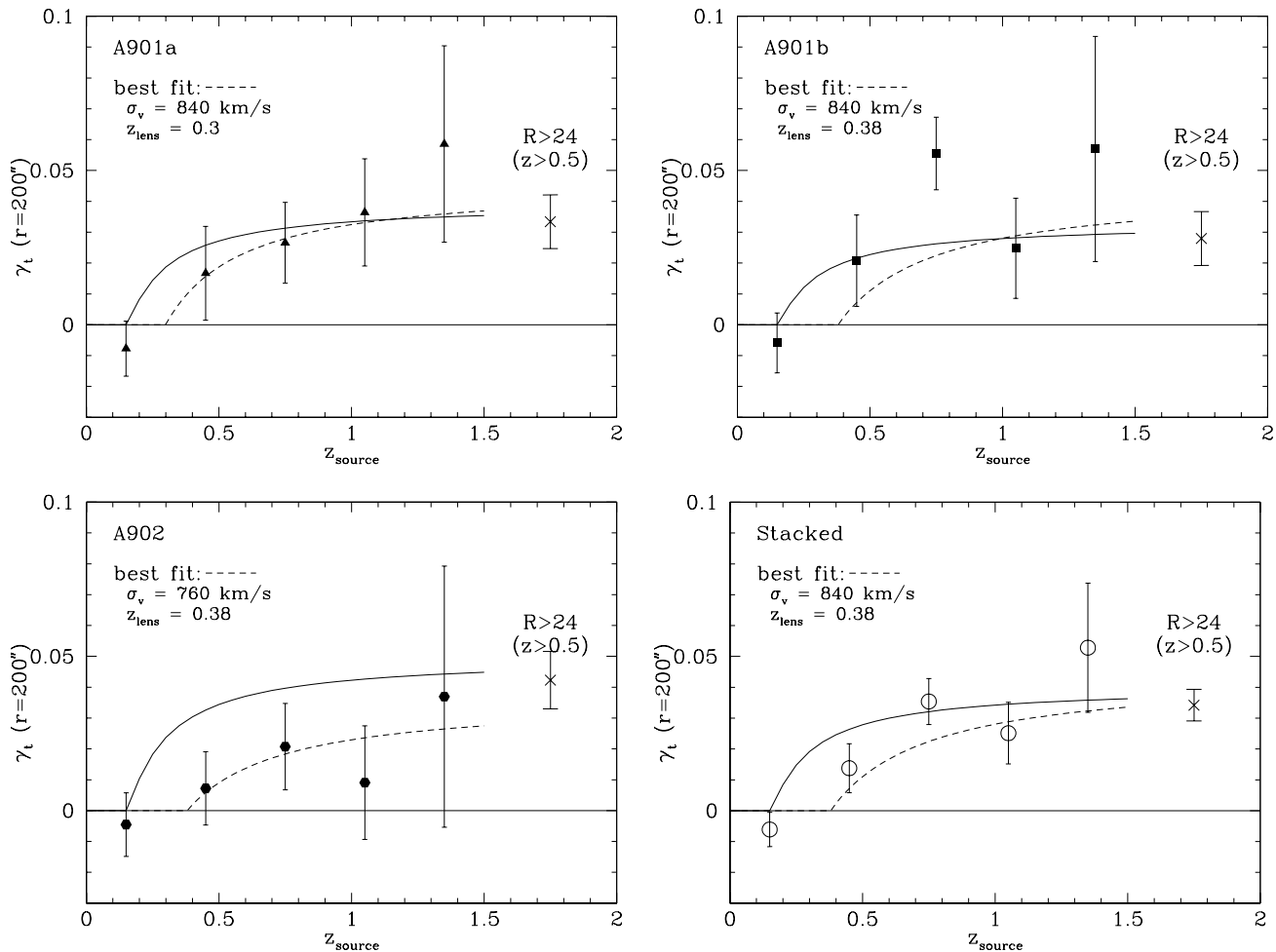,width=\textwidth,angle=0}
 \caption{
Tangential shear as a function of redshift. Each point represents
the amplitude of the tangential shear from the fit to the SIS
radial profile in each redshift slice from
Fig.~\ref{fig-biggtzplot}, plotted at a fiducial radius of 200
arcsec. The dashed line shows the expected redshift dependence
from the best two-parameter fit ($z_{\rm lens}$, $\sigma_v$) shown
in Fig. \ref{fig-gtfitcont}. The solid curve shows the redshift
dependence for a cluster at fixed redshift, with $z_{\rm
lens}=0.16$, and with the amplitude normalised at $z=1$ to the
$R>24$ point.  } \label{fig-gtfitplot}
\end{figure*}

Figure \ref{fig-gtfitplot} shows the tangential shear,
$\gamma_t(\theta)$, as a function of redshift in the direction of each
cluster, and for the `average' stacked cluster. The shear data points
in this plot, and their uncertainties, are from the fit of the SIS
model to the data in Figure \ref{fig-biggtzplot} in each redshift bin,
and plotted at the centre of each bin. As a demonstration of the
growth of the signal, we choose to plot the amplitude of these fits at
the fiducial radius of $\theta=200''$, although we emphasize that the
fit is to {\em all} of the data points in Figure
\ref{fig-biggtzplot}. We show the amplitude of the fit for each of the
redshift bins, and for the $R>24$ sample with unknown redshifts. In
each figure we also show the best-fit shear model with parameters
given by Figure \ref{fig-gtfitcont} (dotted line) and a model with
fixed cluster redshift of $z=0.16$ and a shear amplitude normalised to
the $R>24$ galaxies at $z=1$ (solid line; note that any similar
normalisation at $z>1$ would yield similar results).

The amplitude of the tangential shear as a function of redshift for
A901a, shown in Figure \ref{fig-gtfitplot} (top left), is in good
agreement with the predicted shear for a $z=0.16$ cluster with
velocity dispersion given by Figure \ref{fig-gtfitcont} (dotted
line). In particular the A901a data shows a clear rise of the
tangential shear signal from zero at the cluster redshift.  There is
also good asymptotic agreement with the $R>24$ galaxies.  A901b and
A902 (Figure \ref{fig-gtfitplot}, top right and bottom left) both show
similar trends of increasing shear, although in the case of A901b with
a high shear signal in the $z_s=0.75$ bin. This can be seen in the
higher fit of the model to the data in Figure 2 in the redshift range
$0.6<z<0.9$, and is mainly due to a high shear at $\theta > 200''$.

Finally, in addition to determining the lens strength and redshift,
Figure \ref{fig-gtfitplot} shows that the converse can also be
achieved: by comparing the shear signal in the $R>24$ bin (which
contains $\sim65\%$ of the galaxies in our shear catalogue) with either
those of known redshift or the tangential shear of the best-fit model
of the stacked clusters, the redshift of the $R>24$ sources can be
constrained to be $z_s>0.5$ at the 1-$\sigma$ level. As many lensing
studies suffer significant errors resulting from the unknown redshift
of the background sources, this result illustrates the potential for
calibrating the redshifts of galaxies beyond the limit of both
spectroscopic and photometric redshift surveys (see also Clowe et al
2003). The example given here is rather weak, as all of the major
clusters are at a single redshift, $z=0.16$ and the shear signal
reaches asympotic values relatively quickly. Ideally one would like to
perform such an analysis with clusters at a range of redshifts,
preferably at higher redshift to better constrain the unknown
population.

\subsection{A two cluster model along the line of sight}

In the previous section we have demonstrated the constraints that
the shear data together with the COMBO-17 photometric redshifts
can place on the mass and redshift of a cluster, assuming that all
the deflecting mass is located in a single plane. However, the
redshift survey indicates that the galaxy distribution in the
volume of observation in this field is not dominated simply by the
$z=0.16$ supercluster structure. In particular, we find that the
photometric redshift survey reveals a second cluster, which we
will term CB1, at higher redshift ($z=0.48$) and located only
$\sim$90 arcsec to the southeast of the brightest cluster galaxy
in A902.  In this section we shall exploit this fortuitous
alignment of two clusters and extend our analysis to fit a more
complex mass model along the line of sight. This provides us with
the opportunity to attempt to detect two co-projected clusters via
their gravitational shear (c.f. Bacon \& Taylor 2003, Hu \& Keeton
2003), and measure their redshifts and velocity dispersions.  In
Section 5 we will show that this model is consistent with the
Wiener filtered gravitational potential map which we will obtain
which also shows a cluster-sized mass concentration at this
redshift.

We perform a $\chi^2$ fit for two colinear SIS profiles in a
four-dimensional space of $\chi^2(z_1,\sigma_{v,1},z_2,\sigma_{v,2})$,
for two cluster velocity dispersions and redshifts, to the shear
pattern and photometric redshifts behind the cluster A902, using the
methods outlined above. The cluster redshifts are constrained so that
$z_2>z_1$.  We find that the global minimum in this four-parameter
space is located at $z_1=0.21$, $ \sigma_{v,1} = 350 \,{\rm km}
s^{-1}$ and $z_2=0.45$, $\sigma_{v,2} =650 \,{\rm km} s^{-1}$.  By way
of comparison, when we apply the same approach to A901a and A901b the
first cluster solution in each case is consistent with the
single-cluster model of the previous section, while the second cluster
solution is pushed to the highest redshift and highest mass end of the
four-parameter space.

\begin{figure}
\centerline{\psfig{figure=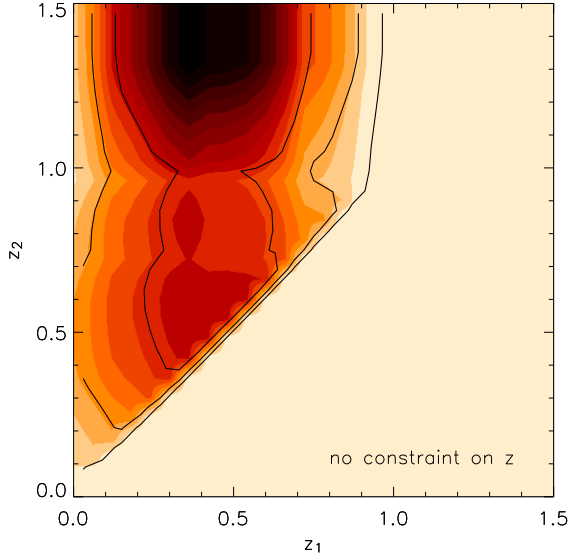,width=\columnwidth}} \caption{
Contours of 68\%, 95\%, and 99.5\% confidence constraints on the
marginalised $\chi^2(z_1,z_2)$ distribution for a two-cluster model
along line of sight through A902 centre. Here we have marginalised
over $\sigma_{v,1}$ and $\sigma_{v,2}$ with $z_1<z_2$.  }
\label{fig-chizz}
\end{figure}

\begin{figure}
\centerline{\psfig{figure=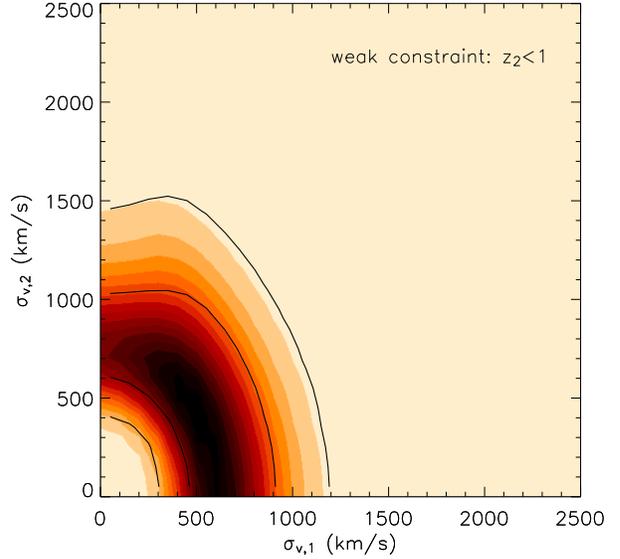,width=\columnwidth}}
\caption{Contours of 68\%, 95\%, and 99.5\% confidence constraints on
the marginalised $\chi^2(\sigma_{v,1},\sigma_{v,2})$ distribution for
a two-cluster model along line of sight through the centre of
A902. The cluster positions have been marginalised over with the weak
constraint that $z_1<z_2<1$. } \label{fig-chimm}
\end{figure}

\begin{figure}
\psfig{figure=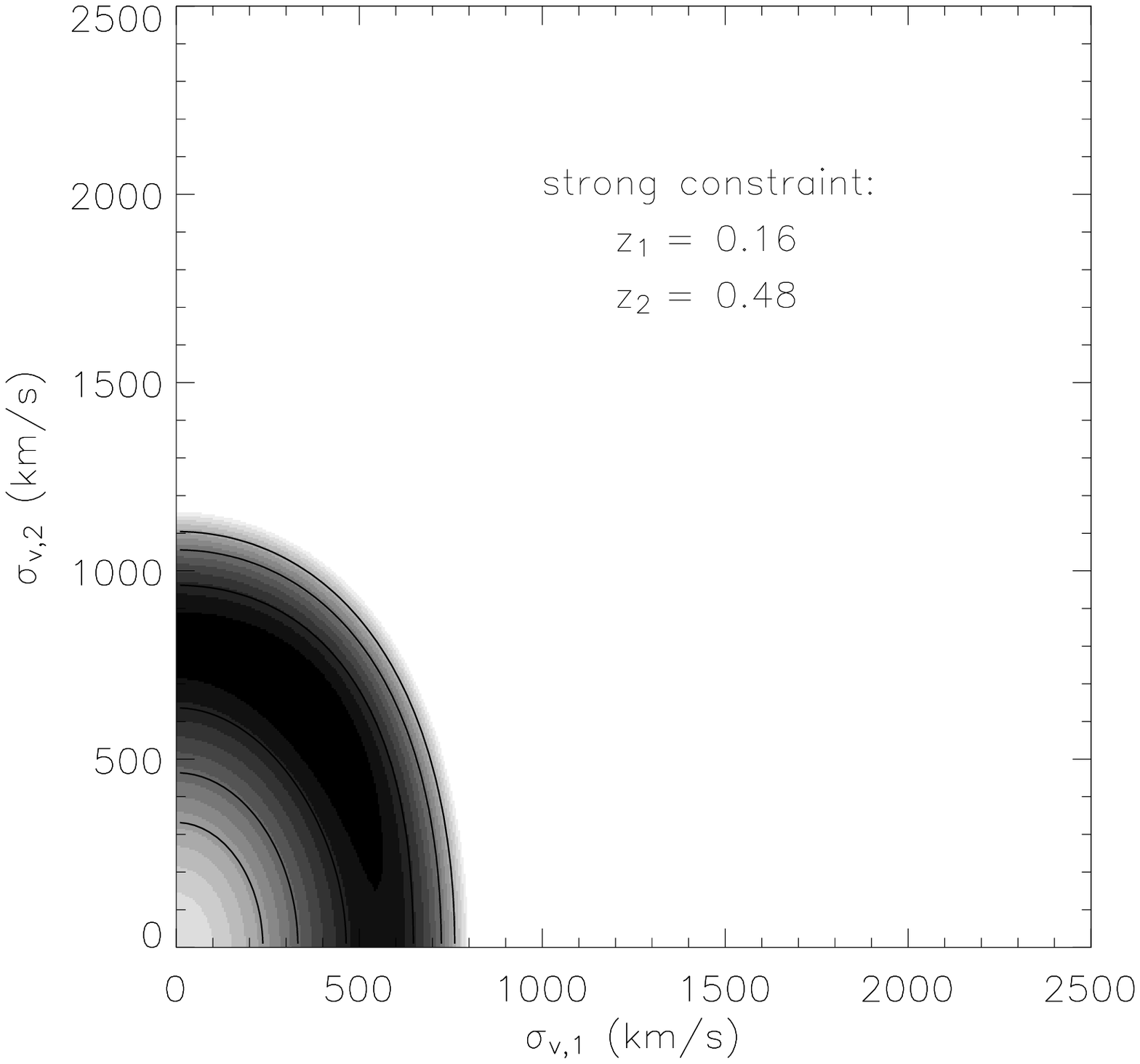,width=\columnwidth}
\hbox{\hspace{0.08\columnwidth}\psfig{figure=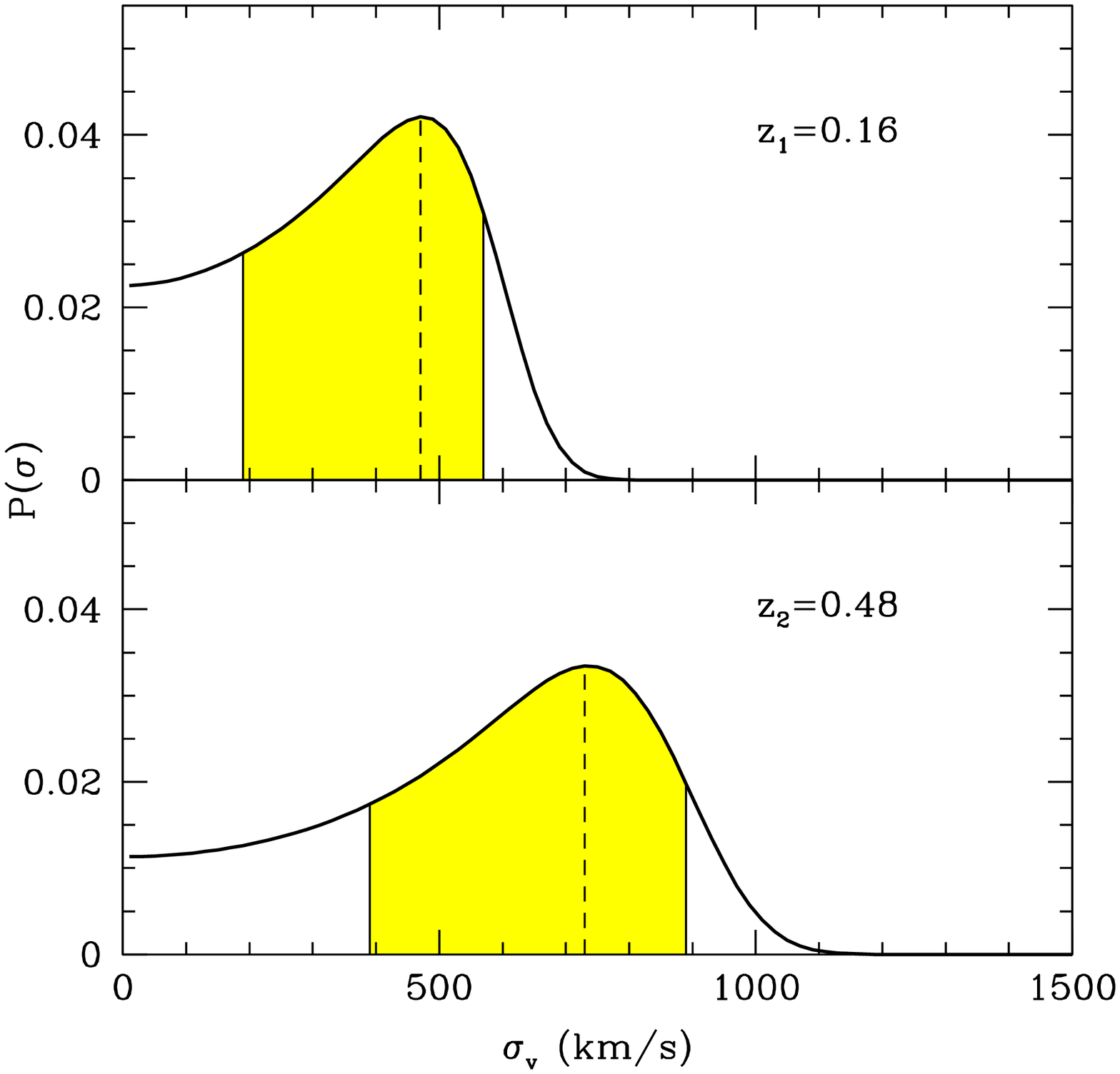,width=0.92\columnwidth}}
\caption{{\em Top:} Contours of 68\%, 95\% and 99.5\% confidence
constraints on the $\chi^2(\sigma_{v,1},\sigma_{v,2})$ distribution
for a two-cluster model along line of sight of A902 centre. Here we
have assumed a strong prior by fixing the known cluster
redshifts. {\em Bottom:} Probability distributions for each cluster
velocity dispersion, marginalizing over the other cluster mass.  The
shaded regions indicate the 68\% confidence regions.}
\label{fig-mmcons}
\end{figure}

Figures \ref{fig-chizz}, \ref{fig-chimm} and \ref{fig-mmcons} show
marginalised projections of the full four-parameter space for the
A902/CB1 clusters. Figure \ref{fig-chizz} shows the 68\%, 95\% and
99.5\% confidence regions for $\chi^2(z_1,z_2)$, after
marginalising over $\sigma_{v,1}$ and $\sigma_{v,2}$ . Since we
require $z_2>z_1$ the lower triangle is excluded. The peak value
for the lower redshift cluster is at $z_1=0.4$. The higher
redshift cluster is best fit at $z_2>1.1$, but with a secondary
minimum visible at $z\sim0.5$. The reason for this can be seen in
the tangential shear as a function of redshift for A902 in Figure
\ref{fig-gtfitplot}. The shear amplitude low in the second bin at
$z=0.5$, so that the fit pushes the low cluster to slightly higher
redshift than its true value at $z=0.16$. The low shear in the
$z=1$ bin also means there is a jump in shear values in higher
redshift bins, causing the fit to infer the second cluster is most
likely here.

Figure \ref{fig-chimm} shows the distribution for
$\chi^2(\sigma_{v,1}, \sigma_{v,2})$ after marginalizing over
redshifts, $(z_1,z_2)$.  Here we have included a weak prior
requiring that $z_1<z_2<1$ to restrict the parameter space to more
reasonable values.  The results, with 68\% confidence limits, are
$\sigma_{v,1}=550_{-300}^{+200} \kms$ and
$\sigma_{v,2}=670_{-550}^{+200} \kms$.  If we drop this weak prior
we find that $\sigma_{v,1}=650_{-200}^{+300} \kms$, while the
second cluster velocity dispersion is unconstrained within the
parameter space.

Finally, we can use our knowledge of the exact positions of A902
and CB1, in this case from the peaks in the 3-D galaxy number
density distribution, to further improve our estimates of the two
masses. We shall refer to this as a strong prior on the cluster
positions, where we perform a $\chi^2$ fit to two cluster velocity
dispersions with fixed redshifts at $z_1=0.16$ and $z_2=0.48$.

Fig.~\ref{fig-mmcons} (top) shows the results for this fit. The
formal minimum lies at a solution with $\sigma_{v_1}=350 \kms$ and
$\sigma_{v_2}=670 \kms$.  The total mass of the two-cluster system
is well-constrained, although again the individual velocity
dispersions are not. However, marginalising over each velocity
dispersion separately (Fig.~\ref{fig-mmcons}, bottom panel) we
find the projected 1-parameter minimum $\chi^2$ values are
$\sigma_{v,1}=470^{+100}_{-280}\kms$ and
$\sigma_{v,2}=730^{+160}_{-340}\kms$. These are then our best
estimates of the masses of A902 and CB1, and are tabulated in
Table 1. This demonstrates that we are able to measure the masses
of two clusters along the same line of sight using shear and
redshift data alone.

Interestingly, even with zero mass in the second cluster
($\sigma_{v,2}=0 \kms$), putting all of the mass in the first
cluster, the most likely value for the velocity dispersion of the
first cluster is $\sigma_{v,1}=570 \pm 120 \kms$. This is slightly
lower than the result of Gray et al (2002) for A902 of
$\sigma_{v}=738^{+244}_{-384} \kms$, although there is a good
overlap with the uncertainties. One possible reason for the change
in the single cluster velocity dispersion of A902, with fixed
redshift, is the different weighting assigned to the galaxy shear
values in these analyses. In Gray et al, where no redshifts were
available, each galaxy was given equal weighting. Here we do have
redshift information. Regardless of the number of galaxies, which
varies from redshift bin to redshift bin, each bin here has
received equal weighting. Hence more weight is given to outliers,
such as the $z=1$ bin in A902, which pulls the mass estimate down.

The lesson we can conclude from this cluster analysis of masses and
positions is the difficulty that a global, parametric fit may
encounter when applied to real data. The positions of clusters in
redshift is especially sensitive to any sharp variations in the
amplitude of the shear. This is due to the fact that additional
clusters are only detected in such a fit by appearance of jumps in the
shear amplitude. Hence our analysis for the {\em positions} of two
projected cluster along the line of sight in the direction of A902 is
not conclusive and so the overall analysis is improved by the addition
of redshift information for the lensing galaxies. However, the
determination of the velocity dispersions is better, so long as we
enforce a weak or strong prior on the cluster positions. While this
demonstrates that some useful constraints on the parametric cluster
masses can be extracted, even with current shear and photometric data,
the problems encountered in this analysis suggest that a
non-parametric approach to cluster finding and characterisation may
also be worth investigation.  In the next section we turn to
reconstructing the full 3-D dark matter gravitational potential from
our data.

\section{Reconstructing the 3-D Dark Matter Potential}

So far, we have used photometric redshifts to interpret our shear
estimates as a 3-D shear field, and found we can constrain the
redshifts and masses of clusters, and even possibly find evidence
for a more complex line-of-sight density distribution. However, as
we described in Section 2, we are now in a position to fully
reconstruct the 3-D gravitational potential using a Kaiser-Squires
(1993) inversion followed by the Taylor (2001) potential
reconstruction method. In this section, we will use these methods
to calculate the gravitational potential for the A901/2 volume of
space. This comoving volume is approximately $\theta^2 r^3/3=
3\times10^{5}[\Mpc]^3$.

As discussed in Bacon \& Taylor (2003), the intrinsic
ellipticities of galaxies, $\sigma_\gamma$, create the major
source of noise upon the gravitational potential reconstruction.
In order to partially overcome this uncertainty, we average the
shear estimators for many galaxies in a cell, chosen to have size
$1.5'\times 1.5' \times \Delta z$ with redshift bin width $\Delta
z = 0.05$; further smoothing can be applied if necessary at a
later stage. We examine galaxies with $0<z<1$; galaxies with
higher redshifts or no redshift assigned are all included in the
final $z=1$ bin.  The full grid size is therefore $20\times 20
\times 20$ cells.

\subsection{The 3-D lensing potential}
In order to calculate the gravitational potential, we must first
determine the lensing potential from the shear field according to
equation (\ref{kaiser-squires}). We estimate this by taking the
Fourier transform of the shear field and using the optimal
weighting of Kaiser \& Squires (1992) to find the Fourier
transform of the lensing potential for each redshift slice;
\begin{equation}
\widehat{\phi}(\k) = 2 ( \hat{k}_x^2 - \hat{k}_y^2 ) \gamma_1 (\k)
+ 4 \hat{k}_x \hat{k}_y \gamma_2 (\k).
\end{equation}
After an inverse Fourier transform to recover
$\widehat{\phi}(r,r\thetab)$, we calculate the coefficients
$\psi_{mn}(r)$ from equation (\ref{psimn}) and correct the
inversion for mean, gradient and paraboloid terms, as described in
Section 2.2, yielding an estimate of $\Delta \phi$ at each slice
in redshift.

We can estimate the uncertainty in a reconstruction of $\Delta
\phi$ from the shot-noise estimates derived by Bacon \& Taylor
(2003), assuming an intrinsic ellipticity of $\sigma_\gamma=0.3$:
 \be
    \lgl \Delta \phi^2 \rgl_{\rm SN} =
    1.7 \times 10^{-15} \left( \frac{n_2}{30/[1']^2}\right)^{-1}
    \left( \frac{\Theta}{1^\circ}\right)^2
    \left( \frac{R^3}{z^2 \Delta z}\right),
    \label{phi_err}
 \ee
where $n_2$ is the projected number density of lensed galaxies,
$R$ is the depth of the survey, and $\Theta$ is the angular radius
of the survey. With the parameters of the COMBO-17 survey this
becomes

 \be
    \Delta \phi_{\rm rms} \approx 5.5 \times 10^{-7} (z/0.1)^{-1}.
 \ee
This is a reasonable estimate of the noise level we find in the
A901/2 field, e.g. $\Delta \phi = 3.45\times10^{-7}$ at $z=0.15$.
However this underestimates the noise by a factor of 3.2 at
$z=0.5$, due to the assumption of a constant space density of
galaxies.

\begin{figure}
\psfig{figure=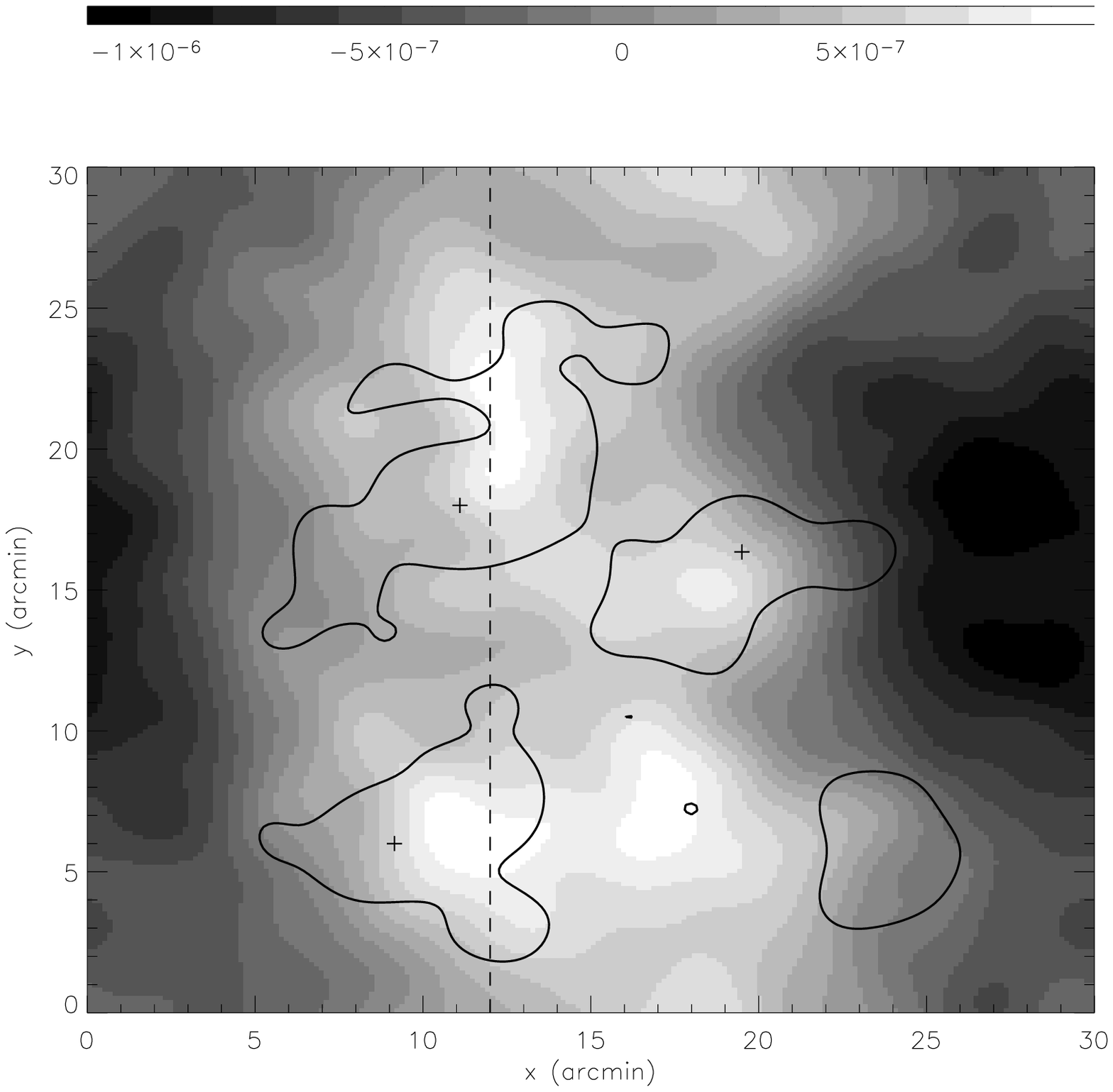,height=70mm}
\psfig{figure=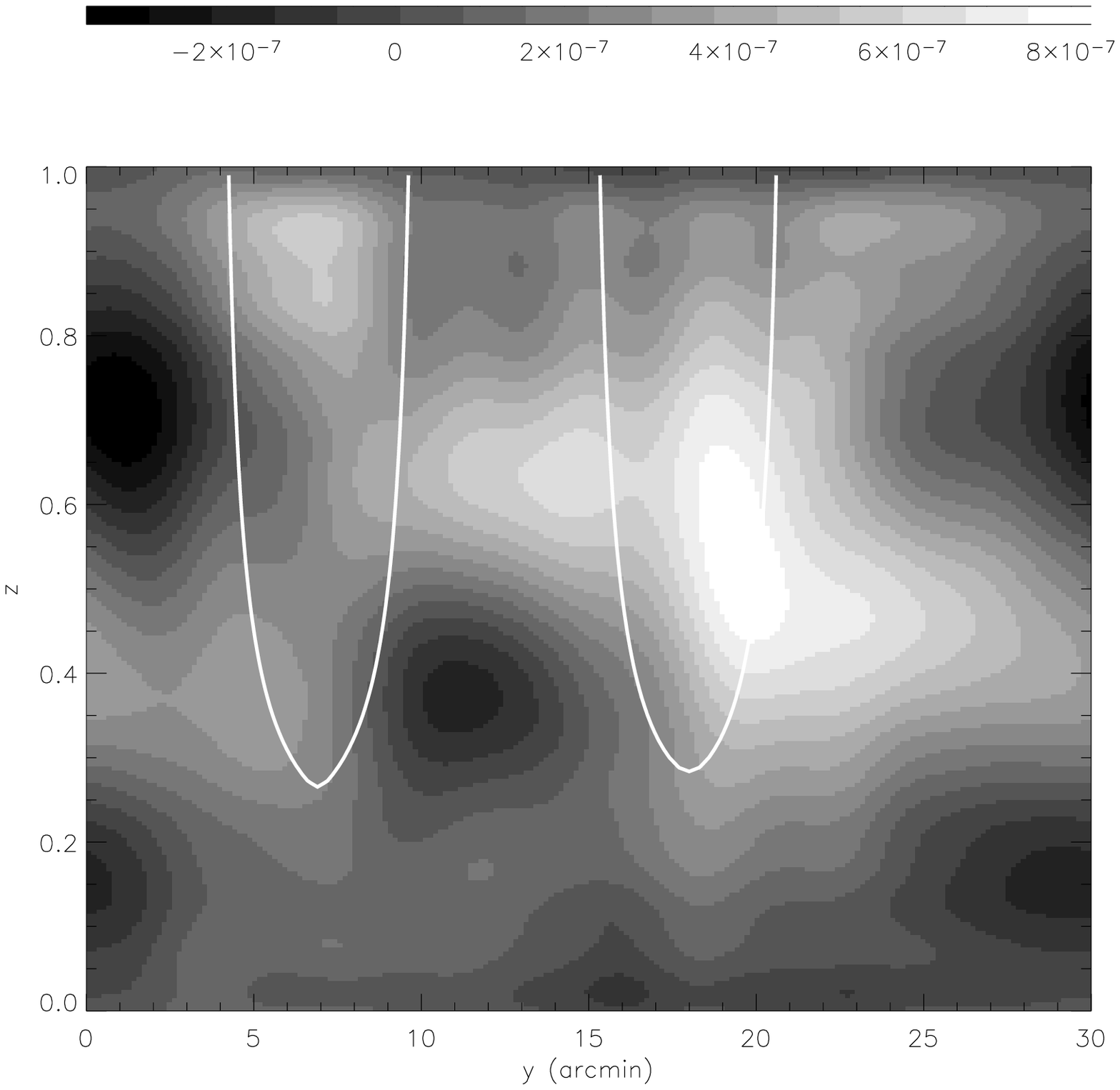,width=75mm,height=70mm}
\caption{Cross-sections of the lensing potential, $\phi$, for the
A901/2 supercluster field. Top panel: $(x,y)$ slice through $z=1$,
by which redshift the lensing potential arising from the clusters
has grown substantially; note the clear detection of the three
cluster signatures (luminosity peaks marked with crosses,
luminosity density contour $L=8 \times 10^{10} L_\odot
[\Mpc]^{-3}$). Bottom panel: $(y,z)$ slice through $x=12'$; this
is a slice through A901a and A902. Note the growth of the lensing
potential signal with redshift behind the two clusters. Solid
lines show theoretical contours of `shark-fin' growth at these
cluster positions).  } \label{fig-phislice}
\end{figure}

\begin{figure}
\psfig{figure=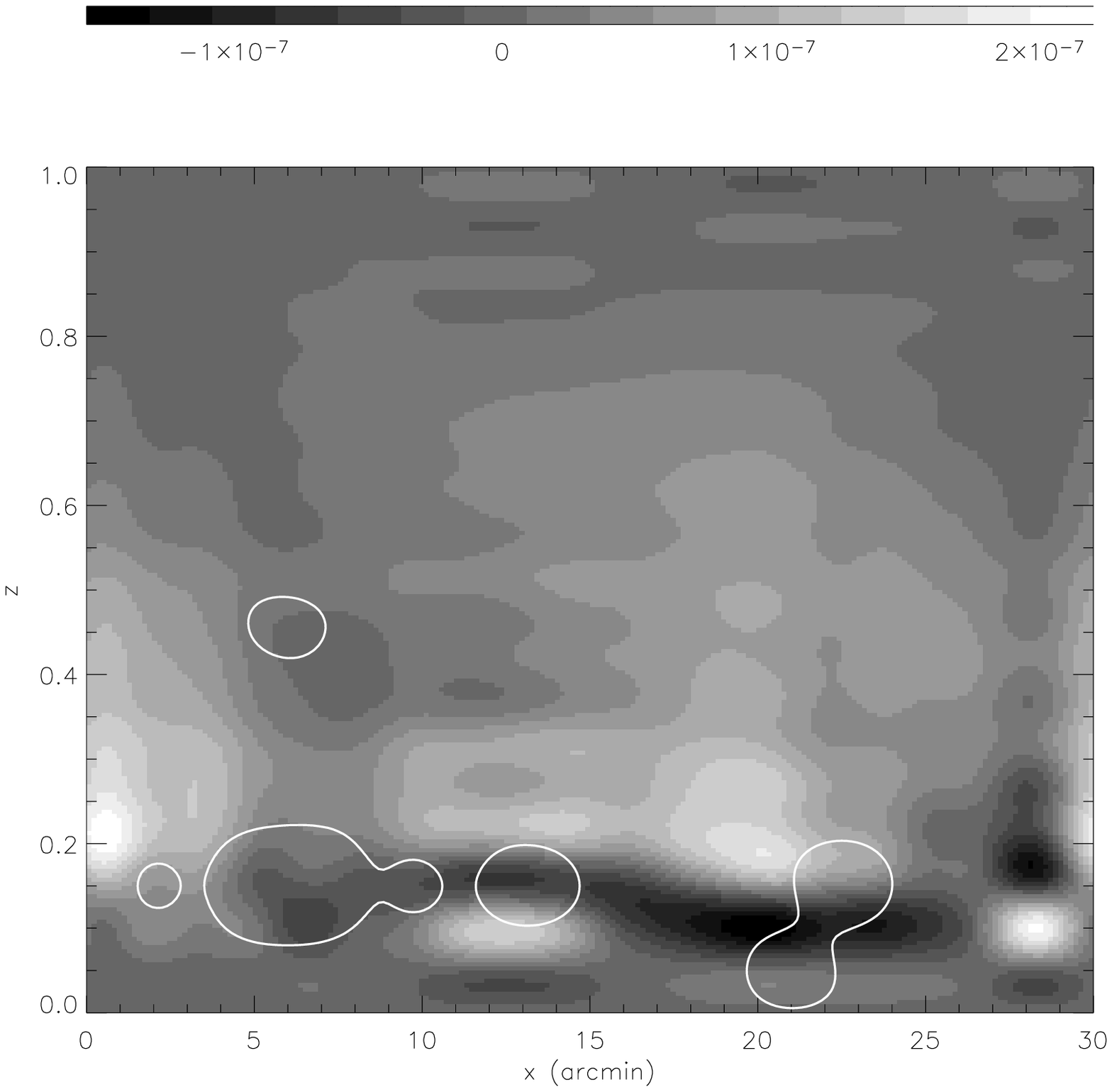,height=70mm} \psfig{figure=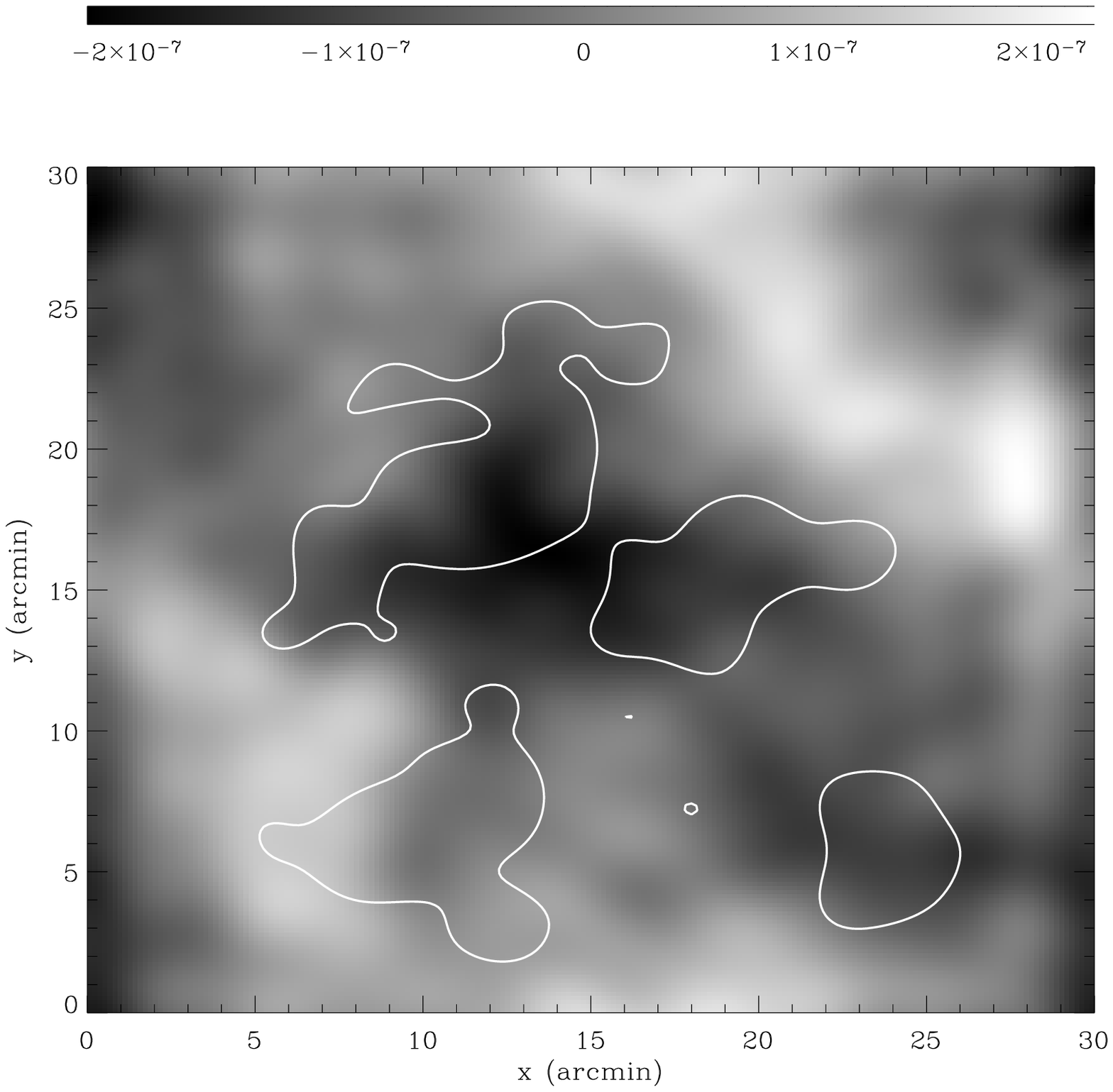,height=70mm}
\psfig{figure=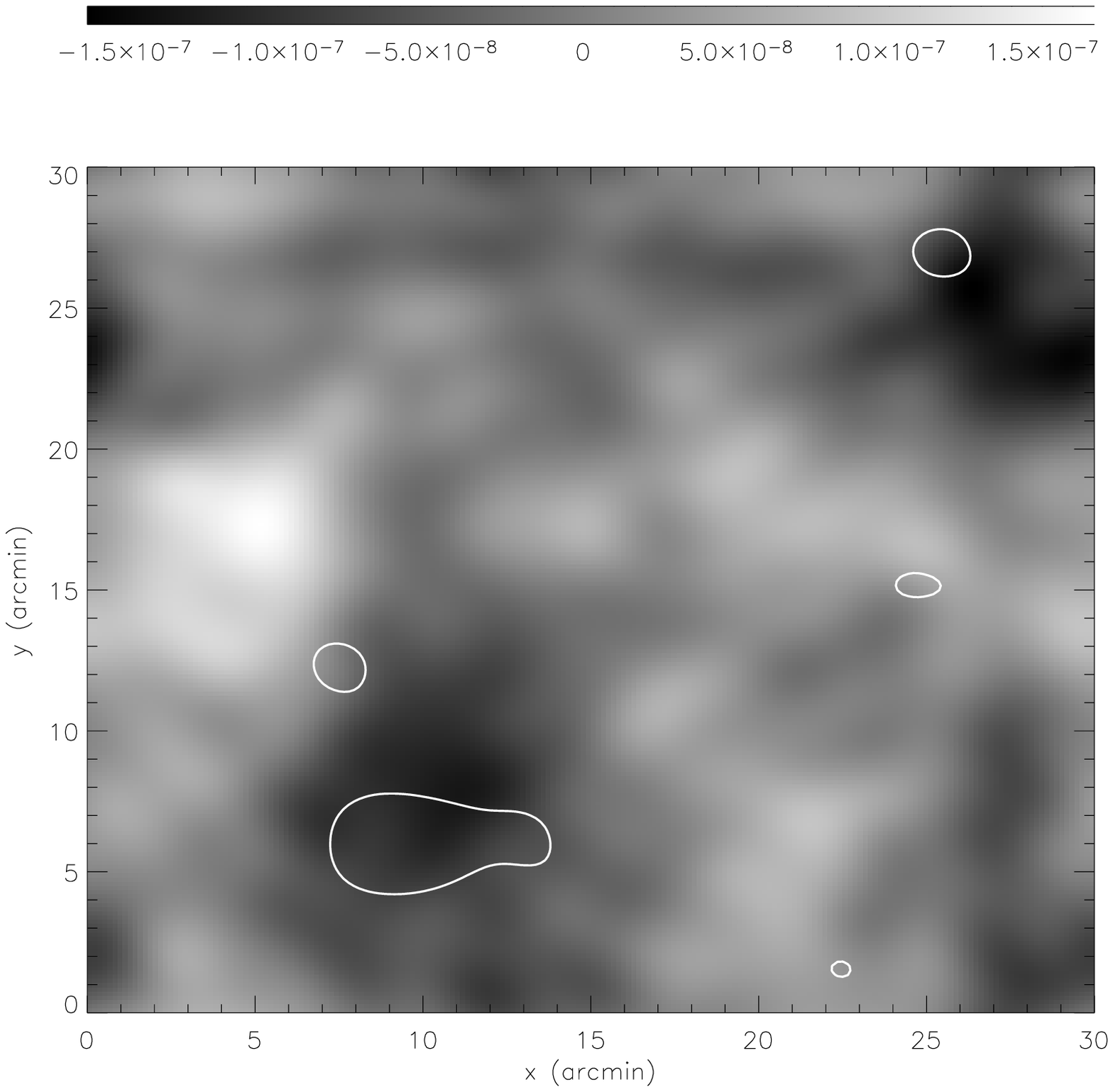,height=70mm} \caption{Cross-sections of the
3-D gravitational potential, $\Phi$, for the A901/2 supercluster
field. Top panel: $(x,z)$ slice through $y=12'$. The gravitational
troughs at $z\simeq0.15$ is associated with the supercluster
(luminosity peaks marked with crosses, luminosity density contour
$L=3 \times 10^{10} L_\odot [\Mpc]^{-3}$). Note also the trough at
$z\simeq0.45$, corresponding to a mass concentration CB1 detected
behind A902. Middle panel: $(x,y)$ slice through the $z=0.15$ to
0.2 slice. We see the troughs associated with the supercluster members
(luminosity density contour $L=8 \times 10^{10} L_\odot
[\Mpc]^{-3}$). Bottom panel: $(x,y)$ slice through $z=0.45$ to
0.5. Note the trough associated with the background cluster CB1
(luminosity density contour $L=2 \times 10^{10} L_\odot
[\Mpc]^{-3}$).  } \label{fig-gravslice}
\end{figure}

\begin{figure}
\psfig{figure=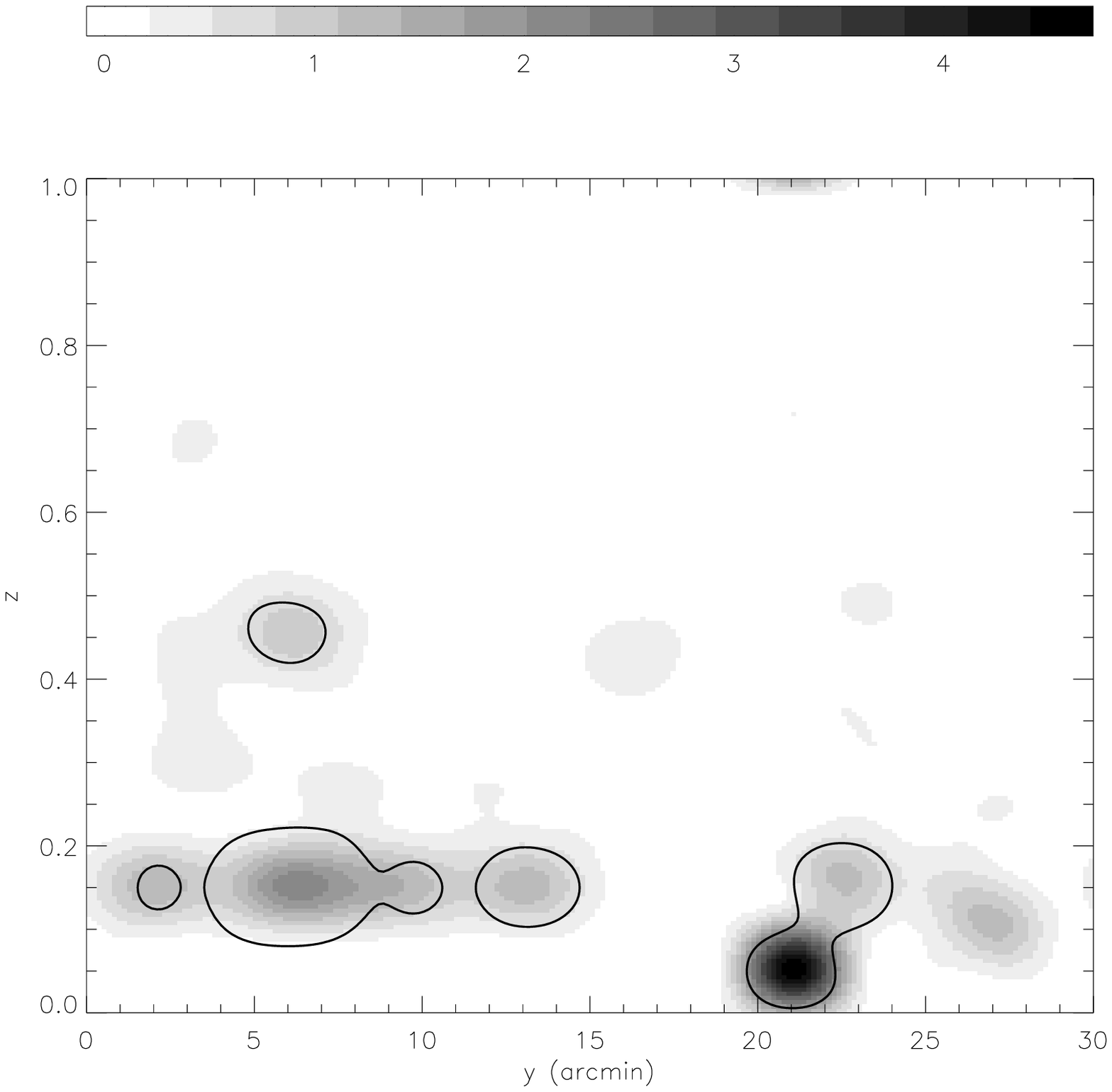,height=70mm} \psfig{figure=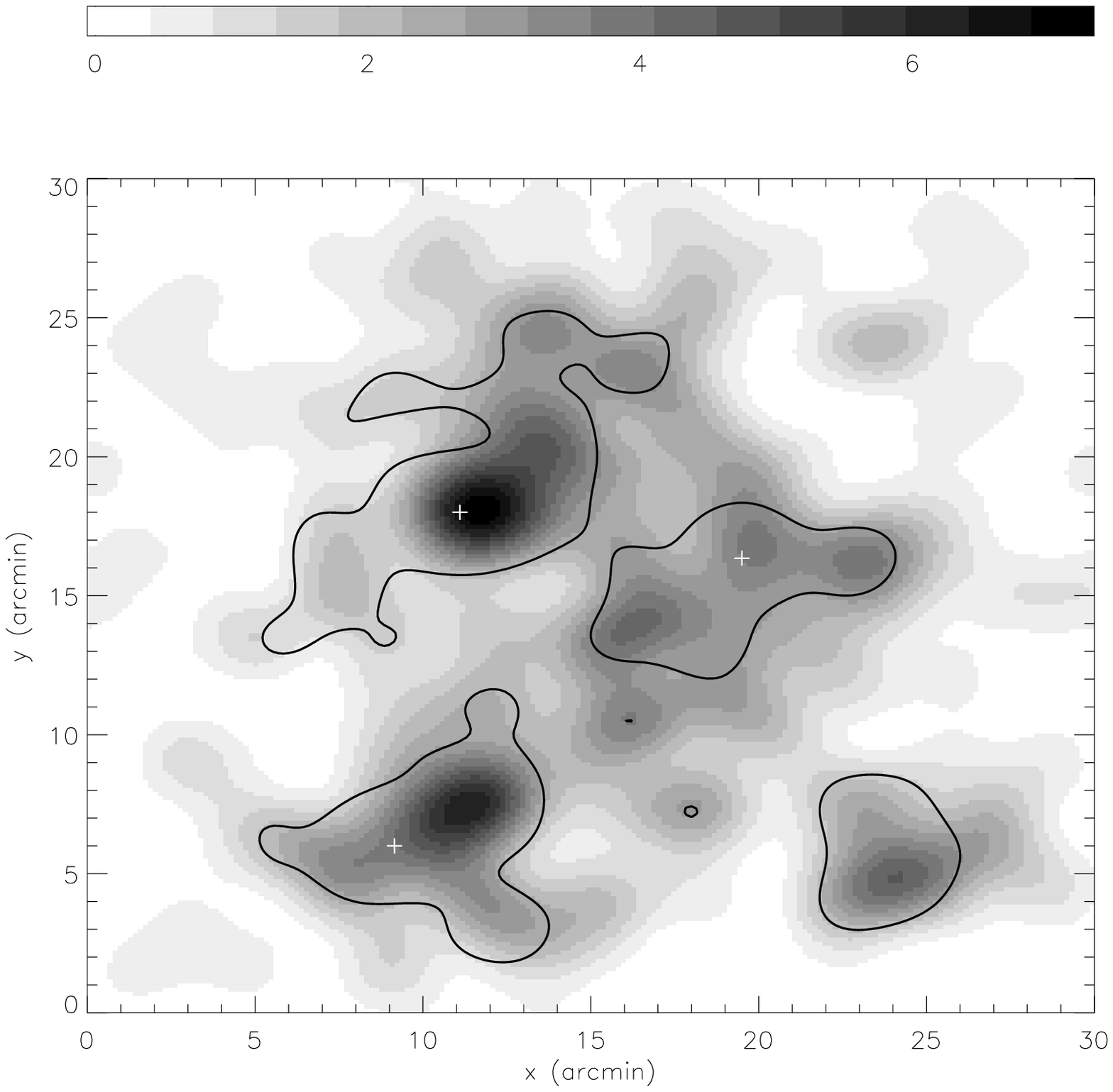,height=70mm}
\psfig{figure=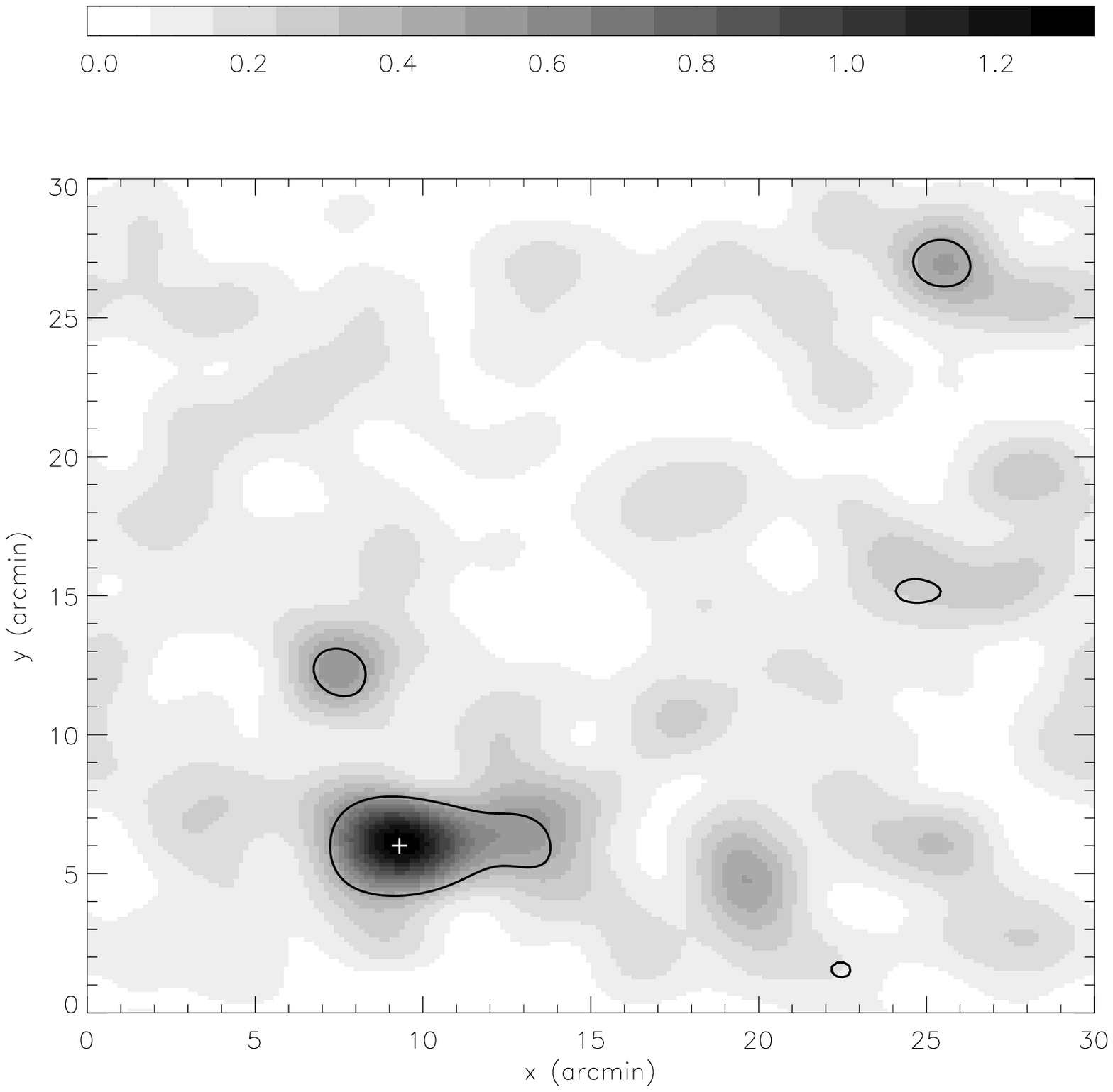,height=70mm} \caption{Cross-sections of the 3-D
galaxy number density $n$ for the A901/2 supercluster field.  Slices
and luminosity contours correspond to those in
Fig.~\ref{fig-gravslice}. Note the overdensities due to the
supercluster at $z=0.16$ and the mass concentration CB1 at $z=0.48$.
} \label{fig-numslice}
\end{figure}

\begin{figure}
\psfig{figure=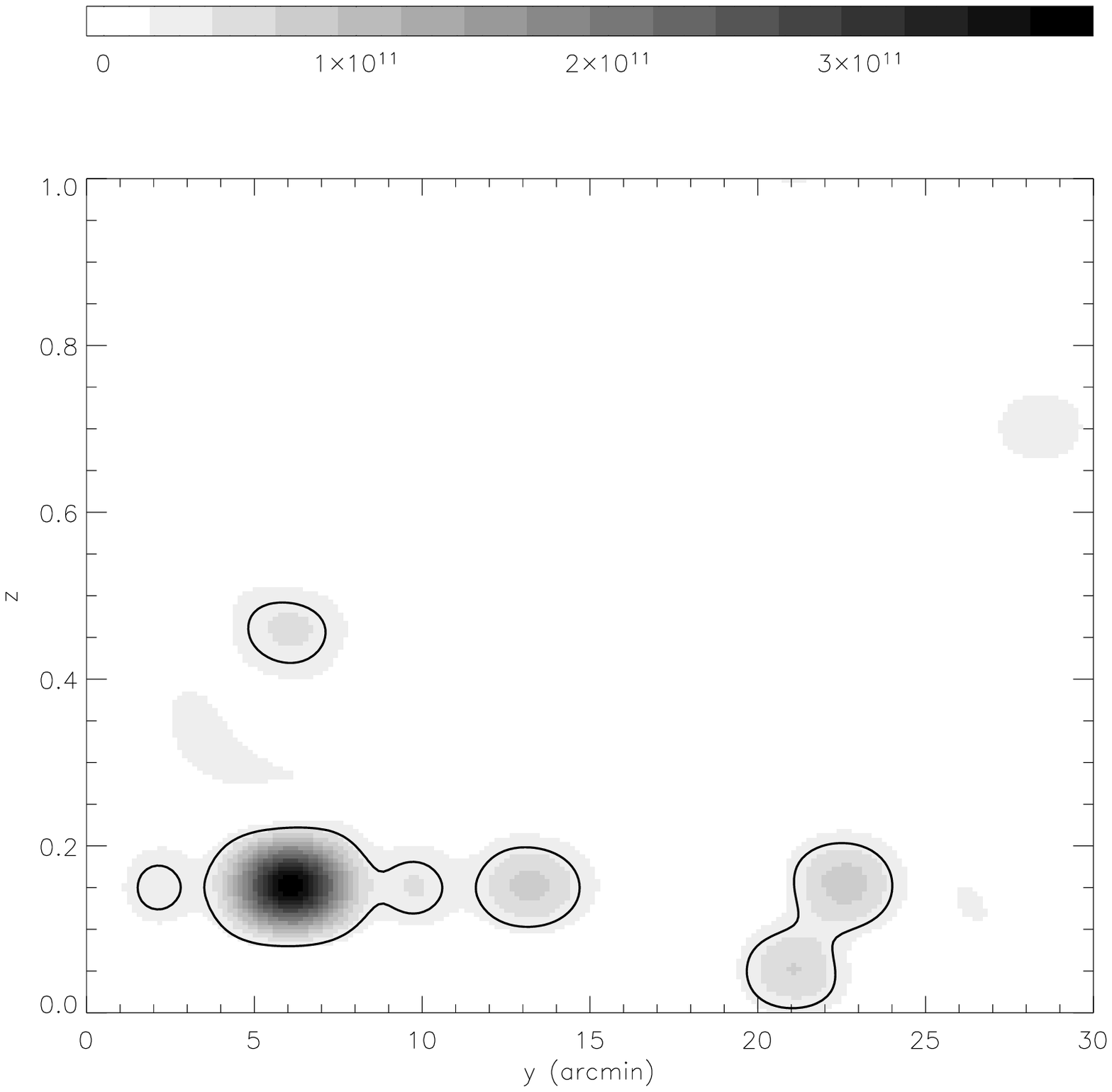,height=70mm} \psfig{figure=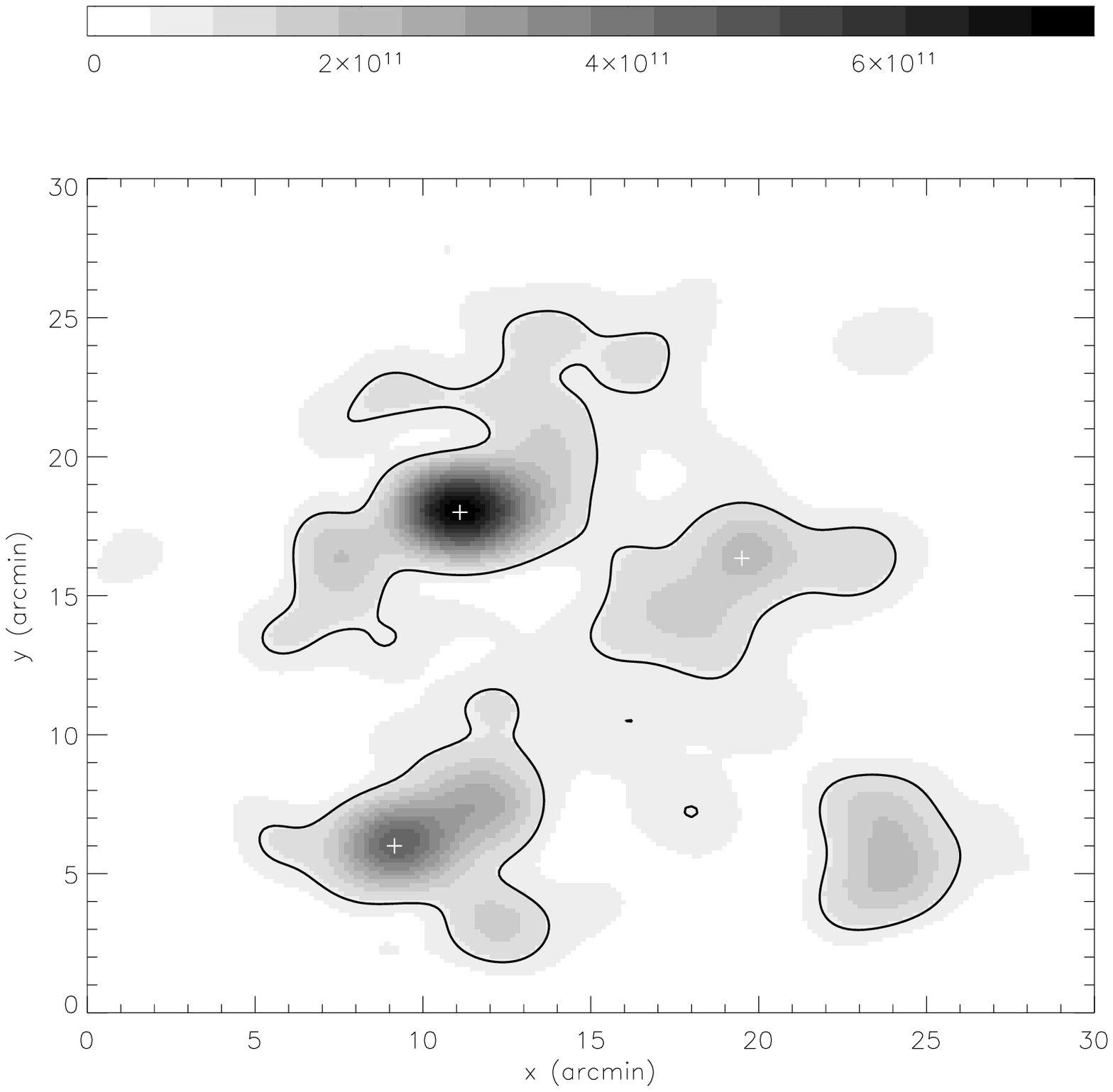,height=70mm}
\psfig{figure=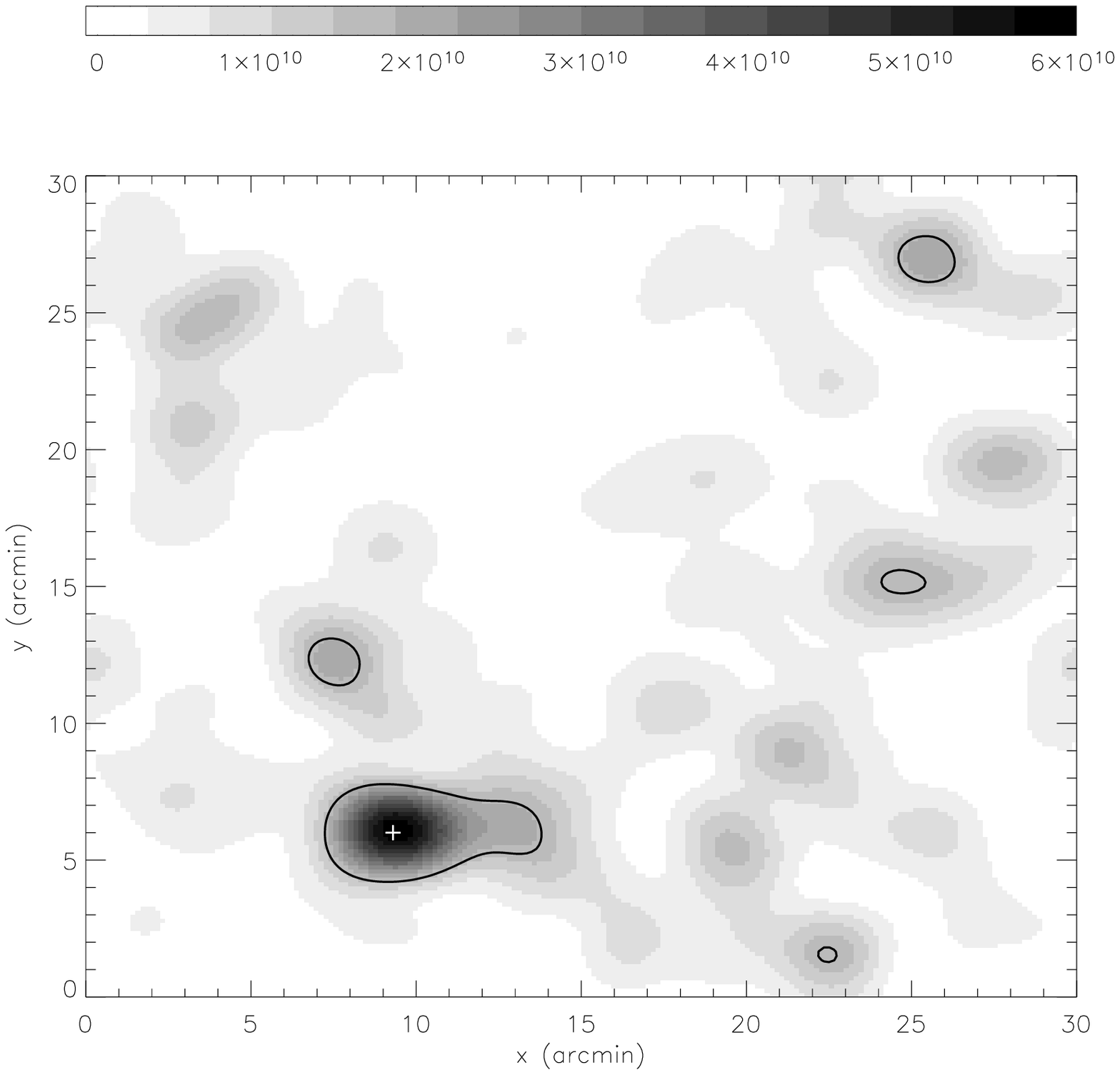,height=70mm} \caption{Cross-sections of the
3-D luminosity density for the A901/2 supercluster field. Slices
correspond to those in Fig.~\ref{fig-gravslice}. Note the
overdensities due to the supercluster at $z=0.16$ and the mass
concentration CB1 at $z=0.48$. } \label{fig-lumslice}
\end{figure}

Figure \ref{fig-phislice} shows two cross-sections of the
reconstructed 3-D gravitational lensing potential, $\Delta \phi$,
field after line-of-sight Gaussian smoothing with a filter scale
of $\Delta z=0.15$. We see in the top panel that the clusters are
individually resolved in the transverse dimensions of the
$\Delta\phi$-field, shown at $z=1$ where the signal-to-noise is
highest. We have overlaid the field with a contour from the 3-D
luminosity density (see Section 5.5) at $L= 8 \times 10^{10}
L_\odot [\Mpc]^{-3}$, indicating a strong correlation between the
lensing potential and luminosity density. Individual 3-D pixels
containing supercluster members have signal-to-noise of 1.5 to 2.1
between $z=0.7$ and $z=1$. We have already seen in Section 4 how
such pixels can be integrated to obtain precise measurements of
the mass of the clusters; here we will continue with moderate
signal-to-noise pixels in order to reconstruct the gravitational
potential.  In the bottom panel, the typical `shark-fin' behaviour
of the lensing potential behind the A901/2 supercluster can be
clearly seen; $\Delta \phi$ grows with redshift behind the
supercluster members.

\subsection{Reconstructing the 3-D gravitational potential}
We can now use equation (\ref{unbiaspot}) to reconstruct the 3-D
gravitational potential, $\Phi$. Following Bacon \& Taylor (2003),
we approximate derivatives in the redshift direction by
second-order differences to reduce error propagation;
\begin{equation}
\partial_z \phi(\r)=
    \frac{\phi(x,y,z-\Delta z)-\phi(x,y,z+\Delta z)}{
    2 \Delta z},
\end{equation}
and
 \be
\partial_{z}^2 \phi(\r)=
    \frac{\phi(x,y,z-\Delta z)+\phi(x,y,z+\Delta z)-
    2\phi(x,y,z)}{(\Delta z)^2},
\ee where $\Delta z$ is the width of a bin in redshift.

The expected signal-to-noise for the potential field can be
estimated from the analytic results of Bacon \& Taylor (2003) who
found that the shot-noise contribution was \be \lgl \Phi^2
\rgl_{\rm SN} = 2.5 \times 10^{-16}
\left(\frac{n_2}{(30/[1']^2)}\right)^{\!\!-1} \!\!  \left(
\frac{\Theta}{1^\circ}\right)^{\!2} \!  \left( \frac{z}{\Delta
z}\right)^{\!4}\!  \left(\frac{R}{\Delta z} \right) , \ee where
$\Delta z=0.05$ is the size of redshift bin.  For the COMBO-17
survey we find \be \Delta \Phi(z) = 8.5 \times 10^{-8} (z/0.1)^2,
\ee compared with an expected signal of $\Phi \sim 10^{-7}$. Since
this implies a signal-to-noise of near unity, we must Wiener
filter to proceed further; we Wiener filter the resulting
$\Phi$-field in the $z$-direction according to equation
(\ref{eqn:wiener}). This completes the process of obtaining a
measurement of the 3-D $\Phi$-field with reasonable
signal-to-noise; we are now able to examine our resulting maps.

\begin{figure*}
\centering
\begin{picture}(200,530)

 \includegraphics{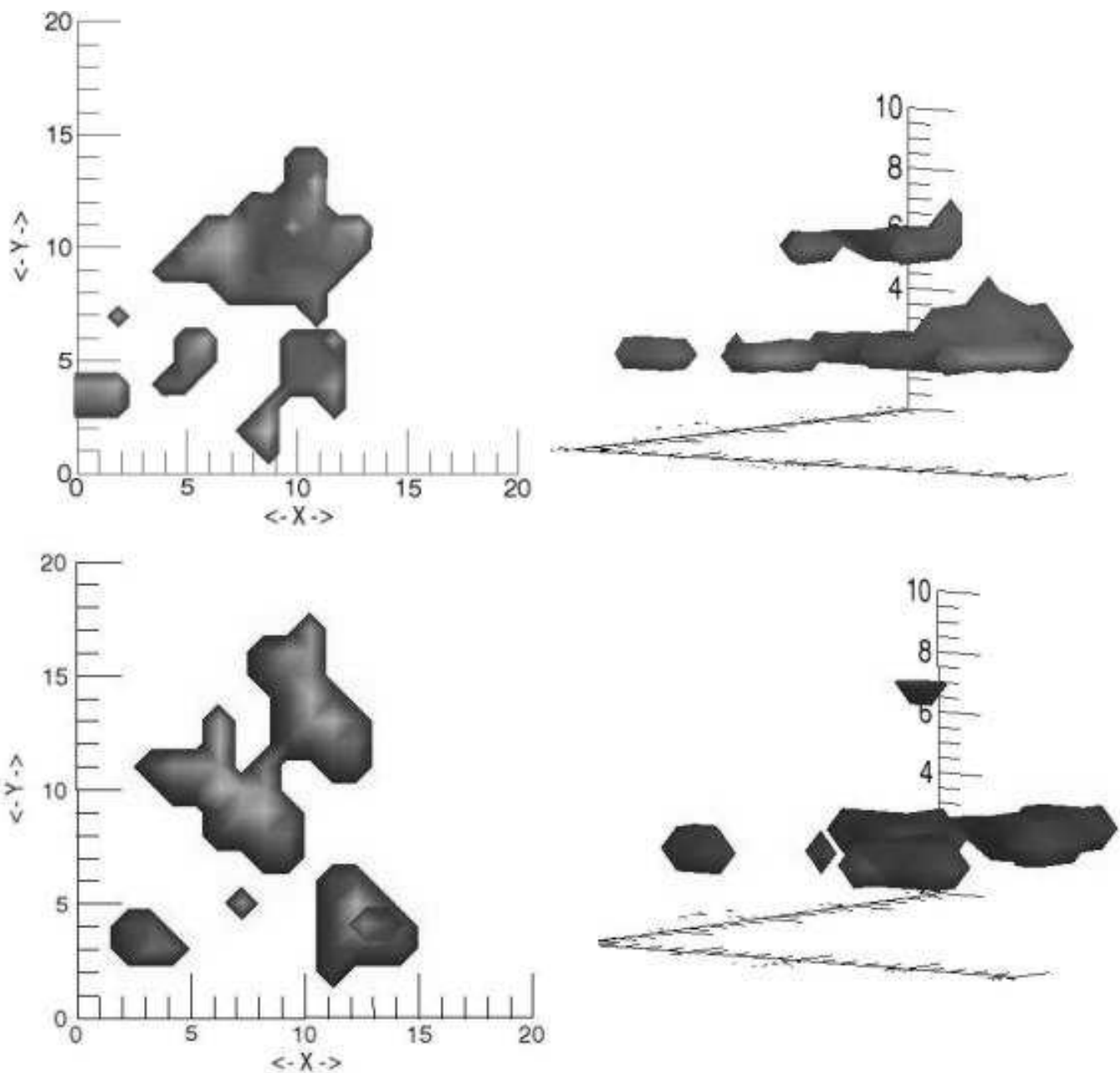}



\end{picture}
\caption{Three-dimensional iso-surface plots of the dark matter
potential and galaxy number density fields. The coordinates of the
map are $(x,y,z)=(\theta_x,\theta_y,z)$, which distorts the map
geometry (note axis are in pixel units, where $\Delta x = \Delta
y=1.5$ arcmins and $\Delta z =0.05$ in redshift). {\bf Upper
panels:} (LHS) The dark matter potential field, seen from
high-redshift looking back to $z=0$  and (RHS) at an oblique
angle. The supercluster A901/2 is seen as a sheet in the potential
field in the lower part of the RHS map. The new cluster CB1 is
clearly seen as an isolated structure in the potential field
behind A902 at $z=0.48$. A902 is at a lower threshold and not
plotted. {\bf Lower panels:} (LHS) The galaxy number density field
for the A901/2 field, in the same projection as the above dark
matter image, and (RHS) in oblique projection. The main
overdensities due to the supercluster are seen as a sheet at low
redshift, while the CB1 cluster is clearly seen at $z=0.48$.
}\label{3D-massmap}
\end{figure*}

\subsection{Mapping the potential}

Figure \ref{fig-gravslice} shows three cross-sections through the
final, Wiener filtered $\Phi$-field. We note several significant
features of this field. The top panel of Figure
\ref{fig-gravslice} shows a slice in the $(y,z)$ plane through
$x=12'$, of the supercluster potential. As for the lensing
potential, we have overlaid this with a contour from the 3D
luminosity density (see Section 5.5) at $L=3 \times 10^{10}
L_\odot [\Mpc]^{-3}$. The gravitational potential well associated
with supercluster A901/2, at $z=0.16$, is clearly recovered with a
peak pixel S/N of 2.7. The cluster centres of A901a and A902 are
also visible clearly visible in this slice.

Beyond the supercluster, the potential field rises as we enter a
void. But beyond that, at $z \simeq 0.48$, there is clearly a
second mass concentration, corresponding to the luminosity peak of
the background cluster CB1 behind A902. This has a peak $\Phi$
pixel S/N of 3.5, and is therefore actually more significant per
pixel than the supercluster itself. This is consequently the first
3-D gravitational potential reconstruction of a cluster behind
another cluster.

The position of CB1 is reasonably well constrained from the galaxy
number distribution (see Section 5.4); $z=0.48\pm0.1$. Bacon \&
Taylor (2003) find that the Wiener filtering only biases the
position of structure by $\Delta z \simeq 0.05$ with ground-based
noise levels, which is consistent with the small offset seen in
the position of the minimum of the CB1 potential well. Hence our
non-parametric map-making approach has succeeded in finding and
deprojecting clusters in the shear field, substantially improving
upon a parametric fit to the raw shear data. However the absolute
depth of the potential well is not so well determined, as the
Wiener filter will bias this by a factor $\sim S/(S+N)$ (equation
\ref{eqn:wiener}). Hence, once a cluster is detected and its
position is determined, a parametric fit is required to estimate
its mass. These mass estimates are presented in Table 1, for both
single and double clusters models, with the redshifts given by the
galaxy number counts.

The middle panel of Fig.~\ref{fig-gravslice} shows a slice in the
$(x,y)$-plane at $z=0.15$ to 0.2, the redshift bin including the
supercluster, overlaid with a contour from the 3-D luminosity
density at $L=8 \times 10^{10} L_\odot [\Mpc]^{-3}$. We see that
the reconstructed gravitational potential wells correspond well to
the positions of the three component clusters of A901/2. We will
assess this quantitatively in Section 6. In this slice A901a has a
peak pixel S/N of 2.7; A901b also has peak pixel S/N of 1.6, while
A902 has peak pixel S/N of 1.4.

The bottom panel of Fig.~\ref{fig-gravslice} shows a second slice
at $z=0.45$, corresponding to the position of the second
structure, overlaid with a luminosity density contour $L=2 \times
10^{10} L_\odot [\Mpc]^{-3}$.  Again we see that the minimum of
the potential well lies close to that of the background cluster
CB1's luminosity peak. The peak signal-to-noise of CB1's
gravitational potential is 3.5, constituting a secure detection
with our map.

Figure \ref{3D-massmap} (upper panels) shows a 3-D image of the
dark matter potential field of the A901/2 supercluster and CB1.
The coordinates are $(x,y,z)=(\theta_x, \theta_y, z)$, which
introduces some distortion. The left-hand panel shows a view of
the supercluster field seen from high-redshift, looking back in
the direction of the observer. Hence the positions of the
supercluster centres is flipped left-to-right compared with the
middle panels of Figure~\ref{fig-gravslice}. The axis of the 3-D
plot are in pixel units, with scaling $\Delta x = - 1.5$ arcmin,
$\Delta y =1.5$ arcmin, and $\Delta z=0.05$ in redshift. The main
features of the supercluster are clearly identified, with the peak
of the cluster luminosities at A901a at $(x,y)=(12,11)$ (in pixel
units), A901b at $(7,10)$ and A902 at $(13,5)$. Here A902 is below
our threshold and so does not appear. The lower feature, centered
at $(10,5)$ is the background cluster CB1. Two other potential
wells appear at $(5,5)$ and $(1,4)$, which also have galaxy
counterparts. The right-hand panel is an oblique view of the
supercluster, with the $x-y$ axis at the bottom and the vertical
$z$-axis. The main A901/2 supercluster complex is clearly seen as
a sheet mass distribution, although there is some leakage of the
potential field to slightly higher redshift from the main A901a/b
complex. This is most probably due to the radial Wiener filtering
of the field.  The CB1 cluster is clearly seen as a separate
structure at higher redshift, $z=0.48$.

\subsection{The galaxy number density distribution}

We can compare our results on the 3-D dark matter potential field
with the visible matter distribution, traced by the galaxy number
density. To estimate this we use the full galaxy redshift
catalogue for the A901/2 field, i.e. all galaxies with photometric
redshifts, rather than only the galaxies used for our shear
analysis.  This provides a total of 15147 galaxies with
photometric redshifts. The galaxy count for each cell in the 3-D
grid is then determined. Using the results of Wolf et al (2003)
for the COMBO-17 survey, we possess an estimate for each cell of
the incompleteness of the galaxy count. For $z<0.8$, the counts
are 90\% complete. In the analysis below, we correct our galaxy
counts by the incompleteness measure as a function of magnitude,
redshift and spectral type (c.f. Wolf et al 2003). We further
introduce an absolute magnitude threshold of $M_B<-19.5$ in order
to counter the effect of seeing many low-luminosity galaxies at
low redshift.

Figure \ref{fig-numslice} shows the Gaussian-smoothed ($\Delta
\theta= 1.5',\Delta z=0.05$) galaxy number density in the same
volume of space as our gravitational potential. The top panel
again shows a slice in $(y,z)$, and again we see the galaxy excess
due to the A901/2 supercluster at $z=0.16$. There is also a peak
in the number counts at $z=0.48$ at the position of the background
cluster CB1. The middle panel of a slice at $z=0.15$ shows up the
three cluster cores, while the bottom panel slice at $z=0.45$
shows CB1 as the most significant galaxy number density at that
redshift. Thus we see that one approach to studying significant
mass concentrations is to detect clusters using number counts, and
compare with gravitational potential maps. Conversely, $\Phi$ maps
may lead to further scrutiny of number density maps to find
visible counterparts to potential wells.

Figure \ref{3D-massmap} (lower panels) again shows a 3-D image, in
$(x,y,z)=(\theta_x, \theta_y, z)$ coordinates, of the iso-number
density surface of the A901/2 supercluster and CB1. The left-hand
panel shows the same viewing angle as the above dark matter
potential, and so the cluster positions are again flipped
left-to-right compared with the middle panel of Figure 11. The
main clusters, A901a, A901b and A902 are all clearly visible at
this threshold, and there is some evidence for another new cluster
again at $(x,y)=(3,3)$ in pixel units. The new cluster CB1 is seen
in projection in front of A902 at $(13,4)$. The right-hand panel
shows the same image rotated to an oblique projection, with the
$x$-$y$ plane at the bottom, with the vertical $z$-axis. Again the
A901/2 supercluster galaxy number density forms a distinct sheet
at low redshift, while the CB1 cluster is again clearly visible at
a higher redshift behind A902. Note that the redshift of CB1 seen
in number density is slightly higher than that of the dark matter
potential well. This slight shift if again probably due to Wiener
filtering of the dark matter potential field.

\subsection{The galaxy luminosity-density distribution}

Finally, we also compare the dark matter potential with the 3-D
luminosity density. In order to do this, we use the absolute
magnitudes, $M_{\rm abs}$, estimated by Wolf et al (2002) for all
galaxies with redshifts. The luminosity, $L= 10^{-0.4(M_{\rm
abs}-M_\odot)}L_\odot$ where $M_\odot$ is the solar magnitude, is
calculated for each galaxy. Using the same grid as before
($3'\times 3' \times [\Delta z = 0.05]$ cells), the luminosity is
summed for each cell. Again we apply the completeness correction
of Wolf et al (2003) as a function of magnitude, redshift and
spectral type, plus an absolute magnitude cutoff of $M_B<-19.5$.

Fig.~\ref{fig-lumslice} shows cross-sections through the 3-D
luminosity density corresponding to the same slices as in
Fig.~\ref{fig-gravslice}, with Gaussian smoothing of $1.5',\Delta
z=0.05$ applied. Again the top panel shows significant peaks in
the luminosity for the supercluster at $z=0.16$ and CB1 at
$z=0.48$. The middle panel shows the three peaks of the A901/2
supercluster clearly. As seen above, there appears to be good
agreement between the mass and light; we will quantify this in
Section 6. The luminosity we measure within a comoving aperture of
$0.5 \Mpc$ for each member cluster is presented in Table 1. We
also estimate the mass-to-light ratio for each of these clusters.

The bottom panel in Fig.~\ref{fig-lumslice} shows a slice in $x$
and $y$ at $z=0.45$, the redshift bin including CB1. Here we find
the expected luminosity peak. Again we present the measured
luminosity, and mass-to-light ratio of CB1 within a comoving
aperture of $0.5 \Mpc$ in Table 1.

\begin{table*}
\begin{tabular}{c|c|c|c|c|c|c|c|}
\hline model & Cluster &  $z$ & $\sigma_v$ & $M(<0.5 \Mpc)$&$L(<0.5\Mpc)$&$M/L$\\
             &         &      & $(\kms)$   &$(\times 10^{13}M_\odot)$&$(\times 10^{11}L_\odot)$ &$(\times M_\odot/L_\odot)$\\
\hline
one cluster&A901a&0.16&$680^{+25}_{-90}$&  $10.8^{+0.8}_{-2.7}$&24.7  & 43.7 & \\
           &A901b&0.16&$600^{+40}_{-85}$&  $8.4^{+1.2}_{-2.2}$&13.5  & 62.2 & \\
           & A902&0.16&$520^{+55}_{-90}$&  $6.3^{+1.4}_{-2.0}$&19.5  & 32.3 & \\
& & & & & & &\\
two clusters&A902&0.16&$470^{+100}_{-280}$&$5.1^{+2.4}_{-4.3}  $&19.5  & 26.2 & \\
            & CB1&0.48&$730^{+160}_{-340}$&$12.0^{+6.0}_{-8.9} $&13.0  & 92.3 & \\
\hline
\end{tabular}
\label{tb:clusters} \caption{Parameters of the four clusters in
the COMBO-17 A901/2 supercluster field. We have assumed $h=0.72$
in these estimates. The cluster velocity dispersions and mass
estimates are from the parameteric fits of Section 4.1 for a
single cluster model, and Section 4.2 for a double cluster model,
with the cluster redshifts fixed at the positions given by the
galaxy number counts.}
\end{table*}

\subsection{Comparison with a 2-D lensing analysis}

We summarize the results of our 3-D lensing analysis in Table 1, which
shows the redshift, velocity dispersion, mass, luminosity and the
cluster mass-to-light ratio within $0.5 \Mpc$ of each cluster centre,
for A901a, A901b, A902 and CB1. The redshifts are taken from the
galaxy number density, while the velocity dispersions, and hence
masses, are estimated from the parametric fits to a single-cluster
model in Section 4.1 for A901a and A901b, and the two-cluster fit of
Section 4.2 for A902 and CB1. Note that in this latter case, the
cluster masses are not independently determined. These may be
compared, to some extent, with the 2-D shear analysis of Gray et al
(2002).

The addition of redshift information has altered the measured
masses of all three clusters relative to the 2-D analysis of Gray
et al (2002); A901a mass has increased by a factor 2.8 in a 3'
aperture, while A901b has decreased its mass by 40\%. But of most
interest is the behaviour of A902: in Gray et al (2002) the
projected mass within an aperture of radius $0.5 \Mpc$ from the
centre of A902 is $16.7 \times 10^{13} M_\odot$ (see again Figure
11, Gray et al 2002). In our 3-D analysis, and after deprojection
with CB1, we find that A902 drops to a mass of
$M=5.1^{+2.4}_{-4.3}\times10^{13}M_\odot$.

While it is tempting to assume that this drop is due to the
deprojection of A902 with CB1 (and the fact that this can be done
is clearly an advantage of the 3-D lensing approach), in this case
the main cause is due to the different weighting of shear data. In
the 2-D case shear was weighted per galaxy, while in the 3-D case
it is weighted per redshift bin. This means the data is
susceptible to bias by outliers. In the case of A902, there is a
clear low outling shear bin at $z=1$ which, though containing
relatively few galaxies, pulls the cluster mass down by a factor
of two compared to the shear at higher redshift.  This effect can
also be seen in the single cluster model, but could be combated
with a larger dataset (e.g. with number densities expected from
space-based observations) and different weighting of redshift
bins.  We conclude that although with 3-D lensing we are able to
measure a large mass for CB1, a cluster of that mass at $z=0.5$ is
not a significant contaminant to the 2-D shear signal of a
foreground structure at $z=0.16$. This can be seen from the fact
that the measured mass of A902 drops only slightly in Table 1 as
we move from a one-cluster to two-cluster analysis.

Luminosities for the clusters have increased substantially (by a
factor $\simeq3$) from the analysis of Gray et al (2002); this is
because the previous study looked at early-type galaxies only, and
had no means of applying a correction for undetected luminosity.

After these changes, and despite the removal of projection
effects, there is still no simple relationship between mass and
light in the A901/2 system, suggesting that it is not in a
dynamically settled state. However, the variation in mass-to-light
in A901/2 is now only by a factor of 2 from cluster to cluster,
rather than the factor of 5 in the previous study.

In this section we have presented 3-D maps of the gravitational
dark matter potential, the smoothed galaxy number counts and the
luminosity density. In order to quantify these distributions
further we now calculate the statistical correlations between
these quantities, and compare then with the predictions from the
halo model.

\section{Mass-Galaxy Correlations}

Now that we have estimated the 3-D gravitational potential in the
A901/2 field we are in a position to examine the auto- and
cross-correlations functions of the 3-D gravitational potential,
the galaxy number density and the galaxy luminosities. In this
Section we describe the procedure used to calculate these
correlation functions and briefly interpret our results. We
compare our results with the halo model developed in Section
\ref{halo}.

We calculate the auto- and cross-correlations using equation
(\ref{corrln}) where the average is taken over all cell pairs in
each slice in redshift bins which are separated by $r_\perp$.
Having calculated the auto- and cross-correlation functions for
each redshift slice, we take the mean of these correlation
functions to measure the overall correlation functions in 3-D (the
physical distance between slices is too large for substantial
correlations). As a simple estimate of the uncertainty, we use
$\sigma_{\rm slice} / \sqrt{N_z}$, where $\sigma_{\rm slice}$ is
the standard deviation of the slice correlation functions and
$N_z$ is the number of slices. This acts as a better measure of
the uncertainty than the standard deviation of correlations for
all pairs, as pairs in a given slice have highly correlated values
for the correlation function.

\subsection{The gravitational potential auto-correlation}

\begin{figure}
\psfig{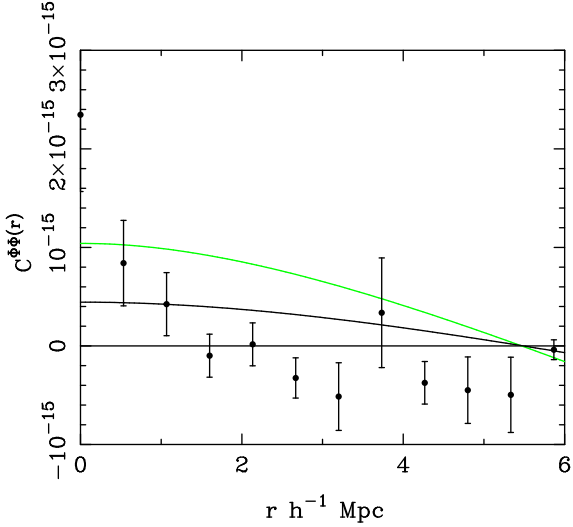} \caption{The auto-correlation
function of the gravitational potential, $\Phi$. Data points are
from a correlation analysis of the A901/2 COMBO-17 field. The dark
solid line is the prediction of the halo model, with a mass
cut-off. The light solid line is the same prediction with no mass
cut-off. }\label{pp}
\end{figure}

The gravitational auto-correlation function, $C^{\Phi \Phi}(r)$,
is shown in Figure \ref{pp}. This shows that the variance at zero
separation is $\lgl \Phi^2 \rgl \approx 2 \times 10^{-15}$,
implying that $\Phi \sim 5\times10^{-8}$, averaged over cells of
size $1.5' \times 1.5'\times \Delta z = 0.05 $. It should be born
in mind that the Wiener filtering in the redshift direction will
affect the amplitude of the reconstructed potential field (Hu \&
Keeton, 2003; Bacon \& Taylor, 2003). Here we have chosen a Wiener
filter which reproduces the masses inferred from the parametric
fits found in Section 4.

The measured potential correlation drops rapidly within $r \sim 2
\Mpc$, then becomes slightly negative before going to zero at $r
\approx 6 \Mpc$. As the mean of the potential field is zero in
each redshift slice, there is a constraint which forces the
integral of the correlation function to be zero over the field.
Hence a positive correlation at small scale must be compensated by
a negative correlation at larger separation.

As discussed in Section \ref{halo}, the halo-model provides an
estimate of the theoretical potential correlations, ensemble
averaged over the whole mass range of collapsed objects. The
lighter solid line in Figure \ref{pp} shows the expected
potential-potential correlation function for all halo masses in a
survey of the size and depth of the A901/2 field, and with the
same pixelisation, and the same integral constraint as our
reconstructed data.

The amplitude at zero separation agrees with the data in the
A901/2 field, although this has been modified by the Wiener filter
to match the cluster velocity dispersions. However the predicted
correlations are higher at larger separation than the data. The
halo model correlation function then passes through zero at a
separation of around $r=5.5 \Mpc$ due to the integral constraint.

In such a finite survey, the high-mass end of the mass function is
not well sampled, as very high-mass peaks appear only rarely in
such a small survey. To account for this, we have also evaluated
the theoretical correlations over a truncated mass range, up to
the mass cut-off of the survey. In the case of the A901/2 field we
truncate the mass range at $M= 10^{13} \,{\rm M}_\odot$,
corresponding to the mass of the main clusters. The dark solid
line is the halo-model expectation value of the potential
correlation function with this mass truncation. With the mass
range truncated the amplitude of the potential correlations drops
to around $\lgl \Phi^2 \rgl \approx 4 \times 10^{-16}$, before
slowly dropping to zero due to the integral constraint.

The shape of the halo-model potential correlations still does not
fit the data from the A901/2 field well. As the gravitational
potential is a long range force, we should expect it to survive to
larger distances, until the integral constraint cuts in. However
here we see that the measured 3-D potential drops rapidly to zero.
There are two possible causes for this. The first is that the
correlation function is dominated by the main clusters in the
field, but that the mass in these clusters is highly concentrated
in the cluster cores. However it is more likely that, while the
correlation function is dominated by the main clusters, the reason
for the shape decline is due to the finite field of view and the
requirement that the mean of the field is zero at each redshift
slice. This will cause us to lose longer-range correlations in the
potential field. Although the halo model does take this integral
constraint into account, it does so averaged over many
realisations of the potential field, and so does not well match
single clusters. In addition the potential reconstruction removes
any long-range potential gradient and quadratic modes in each
redshift slice. However, if we averaged over a number of larger
fields, we would expect the results to converge on the halo-model.

\subsection{The galaxy number density auto-correlation}

\begin{figure}
\psfig{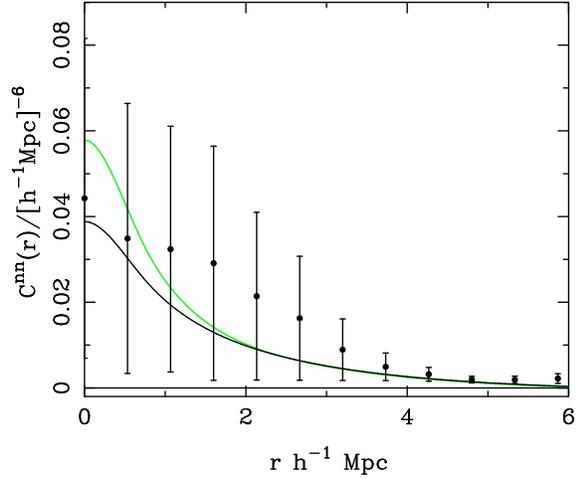} \caption{The auto-correlation of
the number density $n$ from the A901/2 field in the COMBO-17
data-set. Lines are the predictions of the halo model. The lighter
line is for all halo masses, the darker line is for haloes with
mass $< 10^{13} M_\odot$. }\label{nn}
\end{figure}

Figure \ref{nn} shows the auto-correlation function of the galaxy
number density, $C^{nn}(r)$. In this case we have calculated the
correlations between absolute number densities, $n(\r)$, rather
than fractional overdensities, to avoid sensitivity to the mean of
the field, and so there is no integral constraint. Note that we
have truncated the COMBO-17 catalogue at $m_B < -19.5$, to ensure
we are selecting the same type of galaxies as in our model.

At $r=0$ there is a positive correlation, with $C^{nn}\approx 0.04
\, [\Mpc]^{-3}$ due to the galaxies within each cluster. This
falls to zero as we move to separations wider than the clusters.

The halo model does a better job of fitting the data, as the
number density of galaxies is a local property, although the
scatter in the data is large. The lighter line in Figure \ref{nn}
shows the number correlations for all halo masses, while the
darker line shows the correlations for masses $<10^{13} M_\odot$.
Both are in agreement with the trend of the data.

Interestingly the correlations in the galaxy number density seem to
have a longer scale-length than that of the potential. As discussed
above there are two possible explanations for this. One is that the
potential correlations are far shorter than implied by the galaxy
number densities. If this is the case, the dark matter generating the
potential must be very compact. However, it is more likely that this
is again due to the subtraction of the mean potential in the field,
applied to each redshift slice, and the removal of any gradient or
parabolic part of the potential field. These are most likely causing
us to miss part of the larger-scale potential field.

\subsection{The galaxy luminosity auto-correlation function}

\begin{figure}
\psfig{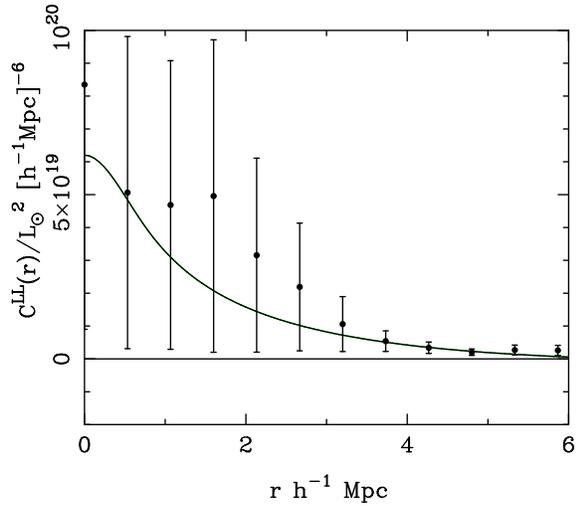} \caption{The auto-correlation
function of galaxy luminosity density, $L$, from the A901/2 field
in COMBO-17. As in the previous figures, the lines are predictions
for the halo model. } \label{ll}
\end{figure}

Figure \ref{ll} shows the galaxy luminosity density projected
auto-correlation function, $C^{LL}(r)$. The COMBO-17 data points
are positively correlated with $C^{LL} \approx 8 \times 10^{19}
L_\odot^2 [ \Mpc]^{-6}$ at zero separation, implying that the
luminosity density is $L \approx 10^{10} L_\odot [\Mpc]^{-3}$ for
galaxies in the A901/2 field with $m_B <-19.5$.

The amplitude at small separations is again dominated by the clusters,
although the scatter is again large. Further out, the signal remains
positive, until around $r=3 \Mpc$, when it dies away to zero. As with
the galaxy number counts, there is no integral constraint on the
luminosity densities.

We have also calculated, for the first time, the prediction of
luminosity densities from the halo model. Here again we see good
overall agreement between the model and data. At small separation
the model peaks at approximately the right amplitude, and then
drops off with increasing distances. Interestingly, the halo model
accounts for the drop-off at large separation very well.

\subsection{The gravitational potential and the galaxy number density}

\begin{figure}
\psfig{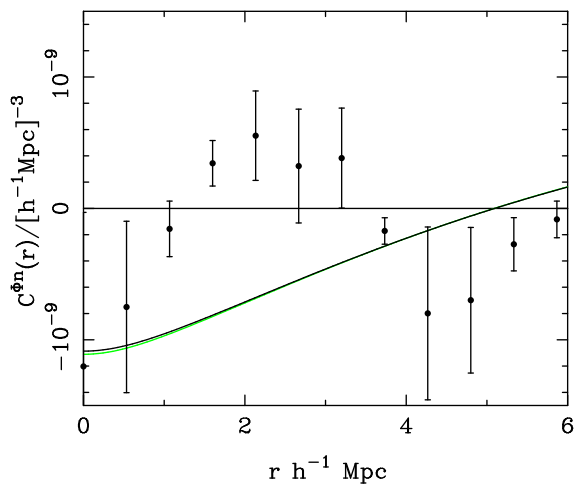} \caption{Cross-correlation of the
gravitational potential $\Phi$ and number density $n$, $C^{\Phi
n}(r)$. The solid lines are, as in previous figures, predictions
from the halo model. }\label{pn}
\end{figure}

Our estimate of the cross-correlation of the gravitational
potential and the galaxy number density is shown in Figure
\ref{pn}. At $r=0$ the signal is dominated by the correlations
between the excess number of galaxies found at the bottom of the
gravitational potential well, and shows a slight anti-correlation.
On larger scales the cross-correlation starts to oscillate due to
the dominance of the supercluster. This oscillation arises as we
move out of the cluster centre and the gravitational potential
increases to a maximum; we find a positive correlation at $r=2.5
\Mpc$ between the potential maximum and the galaxies in the
potential wells. Moving further away, we move into the
gravitational potential of a nearby cluster, and reach a second
anti-correlation at $r=4.5\Mpc$, the typical inter-cluster
distance in the supercluster. On scales larger than $r>5 \Mpc$ we
move out of the supercluster itself.

The predictions of the halo model correlations for all masses
(light line) and for a mass cut-off (dark line) are also shown in
Figure \ref{pn}. This has little effect on the predicted
correlation and the amplitude of the anti-correlation at small $r$
again agrees with our measurement. The halo model does not contain
the oscillations due to the dominance of the main clusters,
indicating again that our sample is too small to converge to
general predictions. The main discrepancy is the positive
correlation at $r \approx 3 \Mpc$. As the predicted galaxy number
density correlations were in fair agreement with the COMBO-17 data
we can conclude that the difference between data and theory lies
with the correlations of the potential field (see Figure 14),
which are not so well matched. However, as noted above, this is
mainly due to the small field size compared with the structures we
are mapping; the imposed zero mean of the gravitational potential
at the supercluster slice has offset the $\Phi-n$ correlation, and
this will be alleviated by using a larger field.

\subsection{The gravitational potential and the galaxy luminosity}

\begin{figure}
\psfig{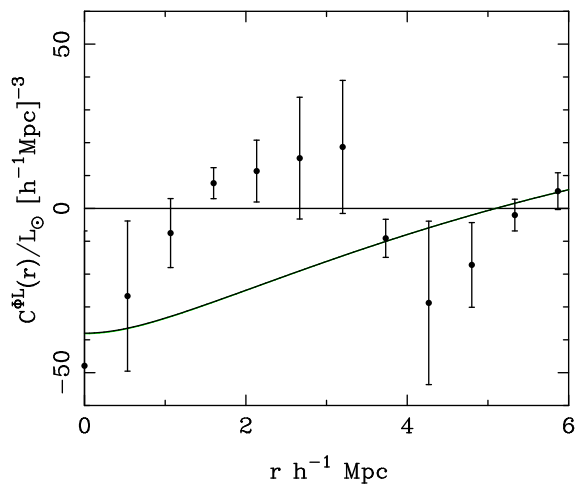} \caption{Cross-correlation of the
gravitational potential, $\Phi$, and luminosity density, $L$,
$C^{\Phi L}(r)$. The solid lines are, as in previous plots,
predictions from the halo model.} \label{pl}
\end{figure}

The $\Phi$-$L$ correlation function, $C^{\Phi L}(r)$, is calculated
for each redshift slice in a similar fashion to the $\Phi$-$n$
correlation function. Again, the mean and variance of the correlation
function found for different redshift slices is used to calculate the
value and error on the overall correlation function. We show the
$C^{\Phi L}$ correlation function in Figure \ref{pl}.

We find only marginal evidence for an anti-correlation between
potential and luminosity at small scales. However we might expect
a high luminosity where there is a deep gravitational potential
well, as suggested by the halo model. As with the potential-number
density correlations, the cross-correlation of potential and
luminosity density again rises to a positive value at around $r
\approx 3 \Mpc$. Again we can identify the discrepancy as being with the
potential field, as we found fair agreement between the halo model
predictions for luminosity-density correlations and the data.
Once more, this will be alleviated by a larger field size.

\section{Conclusions}

In this paper we have for the first time measured the 3-D
gravitational potential of a volume of space from gravitational
lensing. This has been possible by accurately measuring the
gravitational shear on the A901/2 supercluster field of the
COMBO-17 survey, combining these shear measurements with
photometric redshifts, and applying the reconstruction techniques
of Kaiser \& Squires (1993), Taylor (2001) and Bacon \& Taylor
(2003). We have described the COMBO-17 survey, including our
measurements of weak shear (Gray et al 2002) and the photometric
redshift accuracy of the survey, $\Delta z = 0.05$ for $0<z<0.8$.

As a first step towards full 3-D analysis of the A901/2 field, we
have measured the tangential shear around the supercluster members
as a function of redshift. By fitting a singular isothermal sphere
shear model to the shear pattern behind each of these clusters
(c.f. Wittman et al 2001, 2002) we estimated the clusters'
velocity dispersion and redshifts from weak lensing alone. We
found best-fit two-parameter velocity dispersions and redshifts of
$\sigma_v=840^{+100}_{-105} \kms$ and $z=0.30^{+0.10}_{-0.16}$ for
A901a, $\sigma_v=840^{+170}_{-115}\kms$, $z_{\rm
lens}=0.37^{+0.14}_{-0.11}$ for A901b and
$\sigma_v=760^{+220}_{-200}$, $z_{\rm lens}=0.38^{+0.23}_{-0.20}$
for A902. These measurements are consistent with photometrically
determined redshifts of the clusters, $z=0.16$. If we fix the
redshifts of the cluster to the photometric redshift of $z=0.16$,
we find $\sigma_v=680^{+25}_{-90}\kms$ for A901a;
$\sigma_v=600^{+40}_{-85}\kms$ for A901b; and
$\sigma_v=520^{+55}_{-90}\kms$ for A902. We have also compared
with velocity dispersions of the clusters found from a 2-D lensing
analysis (Gray et al 2002) and found good agreement. Any
differences are easily understood as due to the different
weighting of shear data.

Examination of the 3-D number density of galaxies revealed the
existence of a cluster, CB1, at $z=0.48$ behind the A902 cluster.
This prompted us to investigate a two-cluster shear analysis for A902
which provided results consistent with the presence of this second
cluster. Our initial two-cluster analysis failed to improve on the
redshift estimates of the two clusters, due to the difficulty a global
parametric fit to the data has in dealing with outlying shear
values. However, by fixing the positions of the clusters, either with
a weak prior by restricting the redshifts of the clusters to lie below
$z=1$, or with a strong prior of fixing the redshifts to the known
cluster positions, the total mass of the two clusters was well
determined, and each cluster velocity dispersion could be marginalised
over to give $\sigma_{v,1}=470_{-280}^{+100}\kms$ for A902 and
$\sigma_{v,2}=730_{-340}^{+160}\kms$ for CB1.

We have calculated the 3-D lensing potential, $\Delta \phi$, for
this volume of space using the methods of Kaiser \& Squires (1993)
generalised to a 3-D grid. We have then used this lensing
potential to reconstruct the 3-D gravitational potential,
following the reconstruction method of Taylor (2001) and Bacon \&
Taylor (2003). To improve the signal-to-noise in the reconstructed
potential field we Wiener filtered the potential field in the
redshift direction, preserving the properties of the reconstructed
field in the transverse direction.

The recovered gravitational potential was found to include the largest
troughs at the positions of the supercluster members, with a peak
signal-to-noise ratio at the central pixel of $S/N\approx 3$. Hence we
have demonstrated that 3-D dark matter potential mapping is achievable
with currently available shear and photometric redshift data, such as
the COMBO-17 survey.

In addition to the A901/2 supercluster potential, a further
significant ($3\sigma$ per pixel) trough was found at $z=0.48$
behind A902, corresponding to the background cluster CB1 seen in
the 3-D number density and luminosity. This demonstrates that
potential mapping will be a useful tool in the detection of
clusters from lensing data, including clusters behind clusters.
Indeed this suggests an algorithm for the construction of a
mass-selected cluster catalogue, which is free of projection
effects. One first constructs a 3-D potential field from the shear
and photometric data and searches for the largest potential wells.
The positions of these clusters can be accurately determined from
this galaxy number density field, while the cluster masses can be
estimated from parametric fits.

Finally, we have developed the halo model to predict the auto- and
cross-correlation functions of the 3-D potential field, the galaxy
number density field and, for the first time, the galaxy
luminosity-density field, taking into account the geometry of the
survey and pixelisation. We compared these predictions with the
measured auto- and cross-correlations between the gravitational
potential and the galaxy number density, and between the
gravitational potential and the 3-D luminosity distribution in the
COMBO-17 A901/2 field. We found that peaks in the baryonic matter
concentration were strongly correlated to troughs in the
gravitational potential, as one would expect due to baryons
preferentially being found at dark matter concentrations. We also
found reasonable agreement between the amplitudes of the measured
and predicted correlations -- although the statistics of the
A901/2 field are dominated by the four clusters. This introduces
oscillations in the measured correlation functions as we move in
and out of clusters. In particular, the correlation length of the
potential field is shorter than we might have expected from the
clustering of the galaxies. This is most likely due to the removal
of the mean, gradients and parabolic terms in each redshift slice
of the potential field, including the supercluster slice. This
effect will be alleviated with larger area surveys (see Bacon \&
Taylor, 2003, for a further discussion of this effect).

In conclusion, the success of this study in measuring the 3-D
gravitational potential bodes well for future 3-D lensing studies.  We
have seen that 3-D potential mapping will be useful for constructing
mass-selected galaxy cluster catalogues with minimal projection
effects. By extending and comparing with the halo model, we have also
shown that the analysis of haloes within this framework can be
extended to groups and clusters. With advent of better shear and
photometric redshift data, from e.g. the SNAP survey (Massey et al
2003), the prospects look encouraging for 3-D mass mapping large
volumes of the universe, to construct mass-selected cluster
catalogues, to generate statistical probes of dark matter haloes over
a wide range of scales, as a probe of cosmological parameters, and
finally to compare the 3-D dark matter distribution with the
properties of galaxies.

\section*{Acknowledgments}

ANT is supported by a PPARC Advanced Fellowship; DJB and MEG are
supported by a PPARC Postdoctoral Fellowships. CW is supported by
the PPARC Rolling Grant in Observational Cosmology at the
University of Oxford. We thank Alan Heavens and Martin White for
useful discussions.

\label{lastpage}
\end{document}